\documentclass[aps,prd,nofootinbib,amsmath,amssymb,twocolumn,superscriptaddress,10pt]{revtex4}%superscriptaddress,showpacs,

\pdfoutput=1

\usepackage{graphicx}
\usepackage{dcolumn}
\usepackage{bm}
\usepackage{amssymb}
\usepackage{latexsym}
\usepackage{booktabs}
\usepackage{amsmath}
\usepackage{multirow}
\usepackage[colorlinks=true, linkcolor=red, citecolor=blue]{hyperref}

\newcommand{\be}{\begin{equation}}
\newcommand{\ee}{\end{equation}}
\newcommand{\bq}{\begin{eqnarray}}
\newcommand{\eq}{\end{eqnarray}}

\usepackage[usenames,dvipsnames]{xcolor}

\DeclareMathAlphabet\mathbfcal{OMS}{cmsy}{b}{n}
\definecolor{darkgreen}{cmyk}{0.85,0.2,1.00,0.2}
\definecolor{purple}{cmyk}{0.5,1.0,0,0}

\def\barray{\begin{array}}
\def\earray{\end{array}}
\def\be{\begin{equation}}
\def\ee{\end{equation}}
\def\ben{\begin{equation} \nonumber}
\def\een{\end{equation}}
\def\ban{\begin{eqnarray*}}
\def\ean{\end{eqnarray*}}
\def\ba{\begin{eqnarray}}
\def\ea{\end{eqnarray}}
\begin{document}

\title{Real-time cosmology with SKA}
%\title{Cosmological parameter estimation with future redshift drift observations from European Extremely Large Telescope and Square Kilometre Array}
\author{Yan Liu}
\affiliation{Department of Physics, College of Sciences, Northeastern
University, Shenyang 110819, China}
\author{Jing-Fei Zhang}
\affiliation{Department of Physics, College of Sciences, Northeastern
University, Shenyang 110819, China}
\author{Xin Zhang\footnote{Corresponding author}}
\email{zhangxin@mail.neu.edu.cn}
\affiliation{Department of Physics, College of Sciences, Northeastern
University, Shenyang 110819, China}
\affiliation{Ministry of Education's Key Laboratory of Data Analytics and Optimization
for Smart Industry, Northeastern University, Shenyang 110819, China}
\affiliation{Center for High Energy Physics, Peking University, Beijing 100080, China}

\begin{abstract}
In this work, we investigate what role the redshift drift data of Square Kilometre Array (SKA) will play in the cosmological parameter estimation in the future. To test the constraint capability of the redshift drift data of SKA-only, the $\Lambda$CDM model is chosen as a reference model. We find that using the SKA1 mock data, the $\Lambda$CDM model can be loosely constrained, while the model can be well constrained when the SKA2 mock data are used. When the mock data of SKA are combined with the data of the European Extremely Large Telescope (E-ELT), the constraints can be significantly improved almost as good as the data combination of the type Ia supernovae observation (SN), the cosmic microwave background observation (CMB), and the baryon acoustic oscillations observation (BAO). Furthermore, we explore the impact of the redshift drift data of SKA on the basis of SN+CMB+BAO+E-ELT in the $\Lambda$CDM model, the $w$CDM model, the CPL model, and the HDE model. We find that the redshift drift measurement of SKA could help to significantly improve the constraints on dark energy and could break the degeneracy between the cosmological parameters. Therefore, we conclude that redshift-drift observation of SKA would provide a good improvement in the cosmological parameter estimation in the future and has the enormous potential to be one of the most competitive cosmological probes in constraining dark energy.

\end{abstract}
 \maketitle

\section{Introduction}\label{sec:intro}

The accelerated expansion of the universe has been discovered and confirmed by cosmological observations for about twenty years, which is undoubtedly one of the greatest scientific discoveries in the modern cosmology. However, the science behind the cosmic acceleration, i.e., the nature of dark energy, still remains mysterious for us. To measure the physical property of dark energy, one should precisely measure the expansion history of the universe. Currently, the mainstream way is to measure the cosmic distances (luminosity distance or angular diameter distance) and the corresponding redshifts, and to establish a distance-redshift relation, by which constraints on the parameters of dark energy (and other cosmological parameters) can be made. However, a more straightforward way is to directly measure the expansion rate of the universe at different redshifts, although this measurement is more difficult in the observational cosmology.

With the fast advancement in technology over the past several decades, the possibility of measuring the temporal variation of astrophysical observable quantities over a few decades is becoming more and more realistic. This kind of real-time observations can be called the ``{\it real-time cosmology}''. The most typical real-time observable is the {\it redshift drift}, which can give a direct measurement for the expansion rate (namely, the Hubble parameter) of the universe in a specific range of redshift.

The approach of measuring the redshift drift was first proposed by Sandage, who suggested a direct measurement of the redshift variation for the extra--galactic sources \cite{sandage}. At that time, obviously, such a measurement was out of reach with the technological limitation of the day. Then, the method was further improved by Loeb, who suggested a more realistic way of measuring the redshift drift using Lyman-$\alpha$ absorption lines of the distant quasars (QSOs) to detect the redshift variation \cite{Loeb:1998bu}. Loeb concluded that the signal would be detectable when 100 quasars can be observed over 10 years with a 10-meter class telescope. Thus, the method of redshift drift measurement is also referred to as the ``Sandage-Loeb" (SL) test.

Based on SL test, the scheduled European Extremely Large Telescope (E-ELT), a giant 40-meter class optical telescope, is equiped with a high-resolution spectrograph to perform the COsmic Dynamics EXperiment (CODEX). The experiment is designed to detect the SL-test signals by observing the Lyman--$\alpha$ absorption lines within the redshift range of $2\lesssim z\lesssim 5$. The forecast of using the redshift drift from the E-ELT to constrain dark energy models has been extensively discussed; see, e.g., Refs.~\cite{Liske:2008ph,Geng:2014hoa,Geng:2014ypa,Guo:2015gpa,He:2016rvp,Liu:2018kjv,Lazkoz:2017fvx,Zhang:2013mja,Martinelli,Corasaniti:2007bg,Balbi:2007fx,Zhang:2007zga,Zhang:2010im,Geng:2018pxk,Geng:2015ara,Yuan:2013wpa}. It has been shown that the redshift drift in the redshift range of $2< z< 5$ is rather useful to break the parameter degeneracies generated by other observations and thus can play an important role in the cosmological estimation in the future.

Square Kilometre Array (SKA) will soon start construction for the stage of Phase one.  Actually, SKA can also perform the research of real-time cosmology. Instead of detecting the Lyman-$\alpha$ absorption lines of quasar, SKA will measure the spectral drift in the neutral hydrogen (HI) emission signals of galaxies to implement the measurement of redshift drift in the redshift range of $0 < z < 1$. Obviously, the redshift drift data of SKA provide an important supplement to those of E-ELT.

In this work, we will study the real-time cosmology with the redshift drift observation from SKA. We will simulate the redshift drift data of SKA and use these data to constrain cosmological parameters. We have the following aims in this work: (i) We wish to learn what extent the cosmological parameters can be constrained to by using the redshift drift data of SKA-only. (ii) We wish to learn what will happen when the redshift drift data of SKA and E-ELT are combined to perform constraints on cosmological parameters. (iii) We wish to learn what role the redshift drift data of SKA will play in the cosmological estimation in the future.

We will employ several typical and simple dark energy models to perform the analysis of this work. We will consider the $\Lambda$ cold dark matter ($\Lambda$CDM) model in this work, which is the simplest cosmological model and is able to explain the various current cosmological observations quite well. The $w$CDM model is the simplest extension to the $\Lambda$CDM model, in which the equation-of-state (EoS) parameter $w$ of dark energy is assumed to be a constant. The Chevalliear-Polarski-Linder (CPL)~\cite{Chevallier:2000qy,Linder:2002et} model of dark energy is a further extension to the $\Lambda$CDM model, in which the form of $w(a) = w_{0} + w_{a}(1-a)$ with two free parameters $w_0$ and $w_a$ is proposed to describe the cosmological evolution of the EoS of dark energy. We will also consider the holographic dark energy (HDE) model~\cite{Cohen:1998zx} in this work, which is a dynamical dark energy model based on the consideration of quantum effective field theory and holographic principle of quantum gravity~\cite{Li:2004rb}. In the HDE model, the type (quintessence or quintom) and the cosmological evolution of dark energy are solely determined by a dimensionless constant $c$ (note that this is not the speed of light)~\cite{Zhang:2005yz}. For more detailed studies on the HDE model, see e.g. Refs.~\cite{Li:2009jx,Li:2009bn,Li:2004rb,Zhang:2005hs,Zhang:2007sh,Gao:2007ep,Cui:2014sma,Zhang:2014sqa,Zhang:2019ple,Zhang:2006av,Zhang:2006qu,Zhang:2007es,Zhang:2007an,Zhang:2008mb,
Zhang:2009un,Cui:2009ns,Feng:2009hr,Wang:2012uf,Zhang:2014ija,Zhang:2015rha,Zhao:2017urm,Feng:2018yew,Zhang:2015uhk,Wang:2016tsz,Geng:2014hoa,Geng:2015ara,Zhang:2005yz}. In this work, we use these four typical, simple dark energy models, namely, the $\Lambda$CDM, $w$CDM, CPL, and HDE models, as examples to make an analysis for the real-time cosmology.

The structure of this paper is arranged as follows. In Sect.~\ref{Method and data}, we present the analysis method and the observational data used in this work. In Sect.~\ref{sec:Results and Discussions}, we report the constraint results of cosmological parameters and make some relevant discussions. In Sect.~\ref{sec:Conclusion}, the conclusion of this work is given.
\section{Method and data}\label{Method and data}

We will simulate the redshift drift data of SKA, and use these mock data to constrain the cosmological models. We will also simulate the redshift drift data of E-ELT, and make comparison and combination with the data of SKA. In order to check how the redshift drift data of SKA will break the parameter degeneracies generated by other cosmological observations, we will also consider the current mainstream observations in this work.

\subsection{A brief description of the dark energy models}

In this subsection, we will briefly describe the dark energy models employed in the analysis of this work. In a spatially flat universe with a dark energy having an EoS $w(z)$, the form of the Hubble expansion rate is given by the Friedmann equation,
\begin{equation}
\begin{aligned}
  E^2(z)&\equiv\frac{H^{2}(z)}{H^{2}_{0}}=\Omega_{\rm m}(1+z)^{3}+\Omega_{\rm r}(1+z)^{4}\\
 & +(1-\Omega_{\rm m}-\Omega_{\rm r})\exp(3 \int^{z}_{0} \frac{{1+w(z')}}{{1+z'}}dz'),
\end{aligned}
\end{equation}
  where $\Omega_{\rm m}$ and $\Omega_{\rm r}$ correspond to the present-day fractional densities of matter and radiation, respectively. Next, we will directly give the expressions of $E(z)$ for the $\Lambda$CDM, $w$CDM, CPL, and HDE models. Note that since we mainly focus on the evolution of the late universe, in the following we shall neglect the radiation component.
\begin{itemize}
\item $\Lambda$CDM model: Since the cosmological constant $\Lambda$ can explain the various cosmological observations quite well, it has nowadays become the preferred and simplest candidate for dark energy, although it has been suffering the severe theoretical puzzles. The EoS of the cosmological constant is $w=-1$, and thus we have
\begin{equation}
%\begin{aligned}
E^2(z)=\Omega_{\rm{m}}(1+z)^{3}+(1-\Omega_{\rm{m}}).
%\end{aligned}
\end{equation}
\item $w$CDM model: In this model, the EoS of dark energy is assumed to be a constant, i.e., $w={\rm constant}$, and thus it is the simplest case for the dynamical dark energy. For this model, the expression of $E(z)$ is given by
\begin{equation}
 E^2(z)=\Omega_{\rm{m}}(1+z)^{3}+(1-\Omega_{\rm{m}})(1+z)^{3(1+w)}.
\end{equation}
\item CPL model: In this model, the form of the EoS of dark energy $w(a)$ is parameterized as $w(a) = w_{0}+ w_{a}(1 - a)$ with two free parameters $w_{0}$ and $w_{a}$. Thus, we have
\begin{equation}
\begin{aligned}
E^{2}(z)&=\Omega_{\rm{m}}(1+z)^{3}+(1-\Omega_{\rm{m}})\\
&\times(1+z)^{3(1+w_{\rm{0}}+w_{\rm{a}})}\exp\left(-\frac{3w_{\rm{a}}z}{1+z}\right).
 \end{aligned}
\end{equation}
\item HDE model: In this model, the dark energy density is assumed to be of the form $\rho_{{\rm de}}=3c^{2}M^{2}_{\rm{pl}}R_{\rm{eh}}^{-2}$~\cite{Li:2004rb}, where $c$ is a dimensionless parameter, $M_{\rm pl}$ is the reduced Planck mass, and $R_{\rm eh}$ is the future event horizon defined as $R_{\rm{eh}}(t)=ar_{\rm{max}}(t)=a(t)\int_t^\infty{dt'}/{a(t')}$. The evolution of the universe in this model is determined by the following two differential equations,
\begin{equation}
\begin{aligned}
\frac{1}{E(z)}\frac{dE(z)}{dz}=-\frac{\Omega_{\rm{de}}(z)}{1+z}\left(\frac{1}{2}+\frac{\sqrt{\Omega_{\rm{de}}(z)}}{c}-\frac{3}{2\Omega_{\rm{de}}(z)}\right),
\end{aligned}
\end{equation}
\begin{equation}
\begin{aligned}
\frac{d\Omega_{\rm{de}}(z)}{dz}=-\frac{2\Omega_{\rm{de}}(z)(1-\Omega_{\rm{de}}(z))}{1+z}\left(\frac{1}{2}+\frac{\sqrt{\Omega_{\rm{de}}(z)}}{c}\right).
\end{aligned}
\end{equation}
Numerically solving the two differential equations with the initial conditions $E(0)=1$ and $\Omega_{\rm de}(0)=1-\Omega_{\rm m}$ will directly give the evolutions of $E(z)$ and $\Omega_{\rm de}(z)$.
\end{itemize}

\subsection{Current mainstream cosmological observations}

\textbf{SN data:} We use the largest compilation of type Ia supernovae (SN) data in this work, which is named the Pantheon compilation \cite{Scolnic:2017caz}. The Pantheon compilation consists of 1048 SN data, which is composed of the subset of 279 SN data from the Pan-STARRS1 Medium Deep Survey in the redshift range of $0.03 < z < 0.65$ and useful distance estimates of SN from SDSS, SNLS, various low-redshift and HST samples in the redshift range of $0.01 < z < 2.3$. According to the observational point of view, using a modified version of the Tripp formula \cite{Tripp}, in the SALT2 spectral model \cite{Guy}, the distance modulus can be expressed as \cite{Scolnic:2017caz}
\begin{equation}\label{SNu}
{\mu}=m_{\rm{B}}-M+\alpha \times x_{1}-\beta \times c +\Delta_{M}+\Delta_{B},
\end{equation}
where $m_{\rm{B}}$, $x_{1}$, and $c$ represent the log of the overall flux normalization, the light-curve shape parameter, and the color in the light-curve fit of SN, respectively, $M$ repersents the absolute B-band magnitude with $x_{1} = 0$ and $c = 0$ for a fiducial SN, $\alpha$ and $\beta$ are the coefficients of the relation between luminosity and stretch and of the relation between luminosity and color, respectively, $\Delta_{M}$ is the distance correction from the host-galaxy mass of the SN, and $\Delta_{B}$ is the distance correction from predicted biases of simulations.

The luminosity distance $d_{\rm L}$ to a supernova can be given by
\begin{equation}
d_{{\rm L}}(z)=\frac{1+z}{H_{0}} \int_{0}^{z} \frac{dz'}{E(z')},
\end{equation}
where $E(z)= H(z)/H_{0}$. Note that we consider a flat universe throughout this work.
The $\chi^{2}$ function for SN observation is expressed as
\begin{equation}
\chi^{2}_{\rm{SN}}=({\mu}-\mu_{\rm{th}})^{\dagger}C_{\rm SN}^{-1}({\mu}-\mu_{\rm{th}}),
\end{equation}
where $C_{\rm SN}$ is the covariance matrix of the SN observation \cite{Scolnic:2017caz}, and the theoretical distance modulus $\mu_{\rm{th}}$ is given by
\begin{equation}
\mu_{\rm{th}}=5\log_{10}\frac{d_{\rm{L}}}{10\rm{pc}}.
\end{equation}

\textbf{CMB data:} For the cosmic microwave background (CMB) anisotropies data, we use the ``Planck distance priors'' from the Planck 2015 data~\cite{Ade:2015rim}. The distance priors include the shift parameter $R$, the ``acoustic scale'' $\ell_{\rm{A}}$, and the baryon density $\omega_{b}$, defined by
\begin{equation}
R\equiv\sqrt{\Omega_{\rm{m}}H^{2}_{0}}(1+z_{\ast})D_{\rm{A}}(z_{\ast}),
\end{equation}
\begin{equation}
\ell_{\rm{A}}\equiv(1+z_{\ast})\frac{\pi D_{\rm{A}}(z_{\ast})}{r_{\rm{s}}(z_{\ast})},\label{la}
\end{equation}
\begin{equation}
\omega_{b}\equiv \Omega_{b}h^{2},
\end{equation}
where $\Omega_{\rm{m}}$ is the present-day fractional matter density, and $D_{\rm{A}}(z_{\ast})$ denotes the angular diameter distance at $z_{\ast}$ with $z_{\ast}$ being the redshift of the decoupling epoch of photons. In a flat universe, $D_{\rm{A}}$ can be expressed as
\begin{equation}
D_{\rm{A}}(z)=\frac{1}{H_{0}(1+z)}\int_{0}^{z}\frac{dz'}{E(z')},\label{DA}
\end{equation}
and $r_{\rm{s}}(a)$ can be given by
\begin{equation}
r_{\rm{s}}(a)=\frac{1}{\sqrt{3}}\int_{0}^{a}\frac{da'}{a'H(a')\sqrt{1+(3\Omega_{{\rm b}}/4\Omega_{{\rm \gamma}})a'}},\label{rs}
\end{equation}
where $\Omega_{{\rm b}}$ and $\Omega_{{\gamma}}$ are the present-day energy densities of baryons and photons, respectively.
In this work, we adopt $3\Omega_{{\rm b}}/4\Omega_{{\rm \gamma}}=31500\Omega_{\rm{b}}h^{2}(T_{\rm{cmb}}/2.7{\rm K})^{-4}$ and $T_{\rm{cmb}}=2.7255$ K. $z_{\ast}$ can be calculated by the fitting formula \cite{Hu:1995en},
\begin{equation}
z_{\ast}=1048[1+0.00124(\Omega_{{\rm b}}h^{2})^{-0.738}][1+g_{1}(\Omega_{{\rm m}}h^{2})^{g_{2}}],
\end{equation}
where
\begin{equation}
g_{1}=\frac{0.0783(\Omega_{\rm{b}}h^{2})^{-0.238}}{1+39.5(\Omega_{\rm{b}}h^{2})^{-0.76}}, \;  g_{2}=\frac{0.560}{1+21.1(\Omega_{\rm{b}}h^{2})^{1.81}}.
\end{equation}

The three values can be obtained from the Planck TT+LowP data \cite{Ade:2015rim}: $R=1.7488\pm0.0074$, $\ell_{\rm{A}}=301.76\pm0.14$, and $\omega_{b}=0.02228\pm0.00023$. The $\chi^{2}$ function for CMB is
\begin{equation}
\chi^{2}_{\rm{CMB}}=\Delta p_{i}[{\rm Cov}^{-1}_{\rm{CMB}}(p_{i},p_{j})]\Delta p_{j}, \quad \Delta p_{i}=p_{i}^{\rm{th}}-p_{i}^{\rm{obs}},
\end{equation}
where $p_{1}=R$, $p_{2}=\ell_{\rm{A}}$, $p_{3}=\omega_{b}$, and ${\rm Cov}^{-1}_{\rm CMB}$ is the inverse covariance matrix and can be found in Ref.~\cite{Ade:2015rim}.

\textbf{BAO data:} From the baryon acoustic oscillations (BAO) measurements, we can obtain the distance ratio $D_{{\rm V}}(z)/r_{\rm s}(z_{\rm d})$ at the effective redshift. The spherical average gives the expression of $D_{{\rm V}}(z)$,
\begin{equation}
D_{\rm{V}}(z)\equiv\left[D^{2}_{\rm{M}}(z)\frac{z}{H(z)}\right]^{1/3},
\end{equation}
where $D_{\rm{M}}(z)=(1+z)D_{\rm A}(z)$ is the the comoving angular diameter distance \cite{Alam:2016hwk}. $r_{\rm s}(z_{\rm d})$ is the comoving sound horizon size at the redshift $z_{\rm d}$ of the drag epoch and its calculated value can be given by Eq. (\ref{rs}). $z_{\rm{d}}$ is given by the fitting formula~\cite{Hu:1995en},
\begin{equation}
z_{\rm{d}}=\frac{1291(\Omega_{\rm{m}}h^2)^{0.251}}{1+0.659(\Omega_{\rm{m}}h^2)^{0.828}}[1+b_1(\Omega_{\rm{b}}h^2)^{b_2}],
\end{equation}
with
\begin{equation}
\begin{gathered}
b_1=0.313(\Omega_{\rm{m}}h^2)^{-0.419}[1+0.607(\Omega_{\rm{m}}h^2)^{0.674}],\\
b_2=0.238(\Omega_{\rm{m}}h^2)^{0.223}.
\end{gathered}
\end{equation}

We use five BAO data points form the 6dF Galaxy Survey at $z_{\rm eff} = 0.106$ \cite{Beutler}, the SDSS-DR7 at $z_{\rm eff} = 0.15$ \cite{Ross:2014qpa}, and the BOSS-DR12 at $z_{\rm eff} = 0.38$, $z_{\rm eff} = 0.51$, and at $z_{\rm eff} = 0.61$ \cite{Alam:2016hwk}. The data used in this work from various surveys are show in the Table~\ref{bao}.

\begin{table*}[htbp]
\caption{The BAO measurements from the various surveys used in this work. The $r_{\rm s,fid}=147.78$ is the sound horizon for the fiducial model. Note that for the measurements from BOSS-DR12, the first error is the statistical uncertainty, while the second value is the systematic error.}
\label{bao}
\small
\setlength\tabcolsep{3.5pt}
\renewcommand{\arraystretch}{1.5}
\centering
\begin{tabular}{cccc}
\\
\hline\hline
$z$ & Measurement & Experiment & Reference\\
\hline
0.106 & $D_{\rm{V}}(z)/r_{\rm s}(z_{\rm d})=2.976\pm0.133$ & 6dFGS &~\cite{Beutler} \\
0.15 &$D_{\rm{V}}(z)/r_{\rm s}(z_{\rm d})=4.466\pm0.168$ &SDSS-DR7 &~\cite{Ross:2014qpa}\\
0.38 & $D_{\rm{M}}(z)(r_{\rm s,fid}/r_{\rm s}(z_{\rm d}))=1512\pm27\pm14$ &BOSS-DR12 &~\cite{Alam:2016hwk}\\
0.38 & $H(z)(r_{\rm s,fid}/r_{\rm s}(z_{\rm d}))=81.2\pm2.2\pm1.0$ &BOSS-DR12 &~\cite{Alam:2016hwk}\\
0.51 & $D_{\rm{M}}(z)(r_{\rm s,fid}/r_{\rm s}(z_{\rm d}))=1975\pm27\pm14$ &BOSS-DR12 &~\cite{Alam:2016hwk} \\
0.51 & $H(z)(r_{\rm s,fid}/r_{\rm s}(z_{\rm d}))=90.9\pm2.1\pm1.1$ &BOSS-DR12 &~\cite{Alam:2016hwk} \\
0.61 & $D_{\rm{M}}(z)(r_{\rm s,fid}/r_{\rm s}(z_{\rm d}))=2307\pm33\pm17$ &BOSS-DR12 &~\cite{Alam:2016hwk}\\
0.61 & $H(z)(r_{\rm s,fid}/r_{\rm s}(z_{\rm d}))=99.0\pm2.2\pm1.2$ &BOSS-DR12 &~\cite{Alam:2016hwk}\\
\hline\hline
\end{tabular}
\end{table*}

\begin{figure*}
\center{\includegraphics[width=18cm]{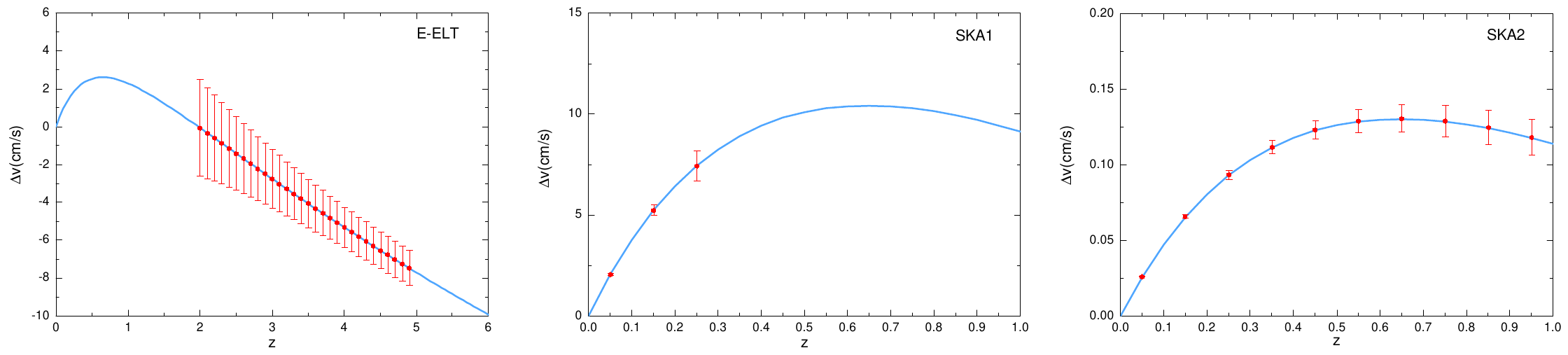}
\caption{\label{mockDV} Curve of $\Delta v$ versus $z$ in the $\Lambda$CDM model. Parameter values are fixed as the best-fit values to SN+CMB+BAO. The error bars on the curves are estimated from E-ELT ({\it left}), SKA1 ({\it middle}) and SKA2 ({\it right}). }}
\end{figure*}

\begin{figure*}[htbp]
\centerline{\includegraphics[width=8cm]{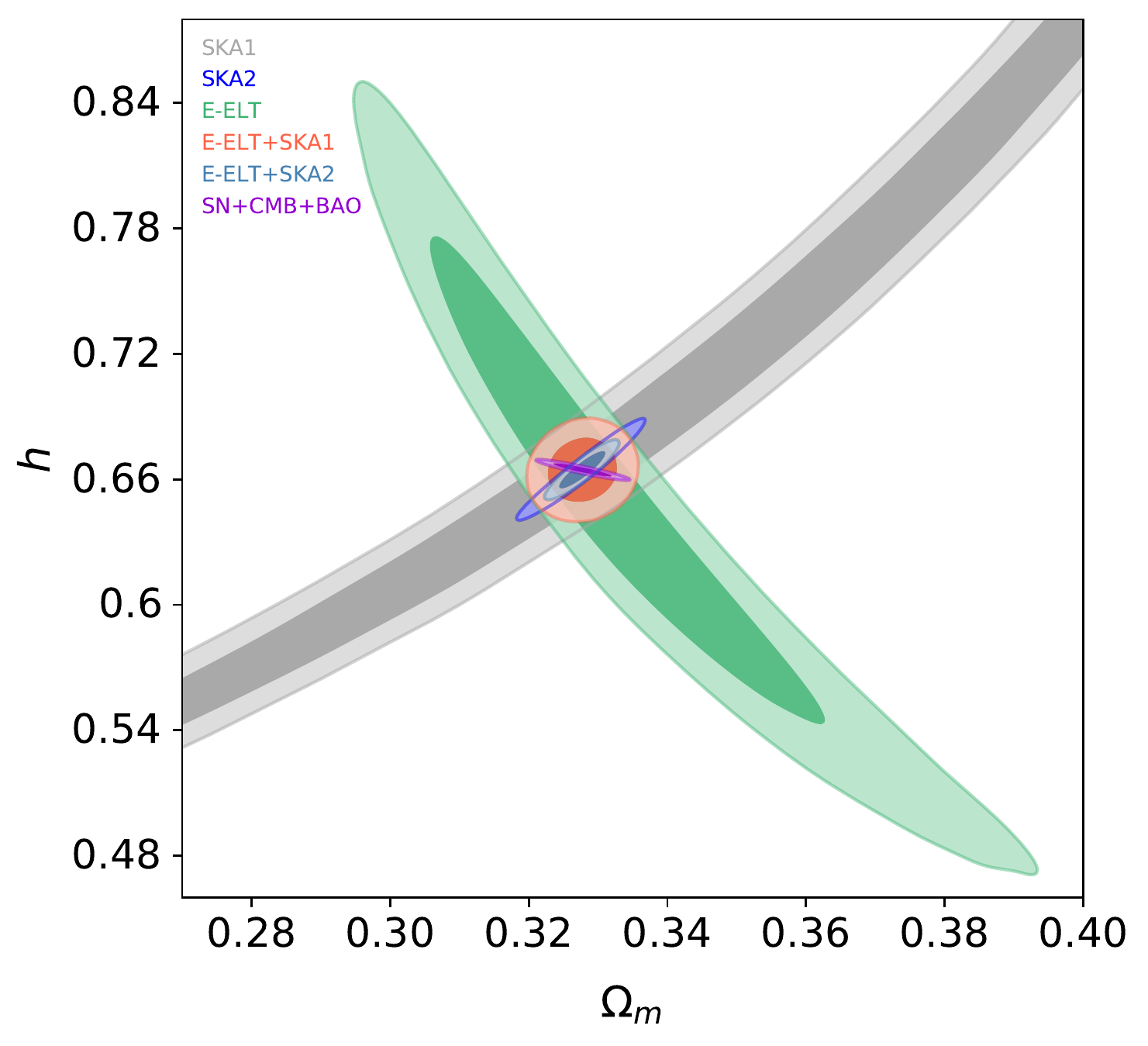}}
\caption{\label{ELTvsSKA} Constraints ($1\sigma$ and 2$\sigma$ CL) on the $\Lambda$CDM model in $\Omega_{\rm m}$--$h$ plane by using the SKA1, SKA2, E-ELT, E-ELT+SKA1, E-ELT+SKA2, and SN+CMB+BAO data.}
\end{figure*}

\begin{figure*}
\centerline{\includegraphics[width=16cm]{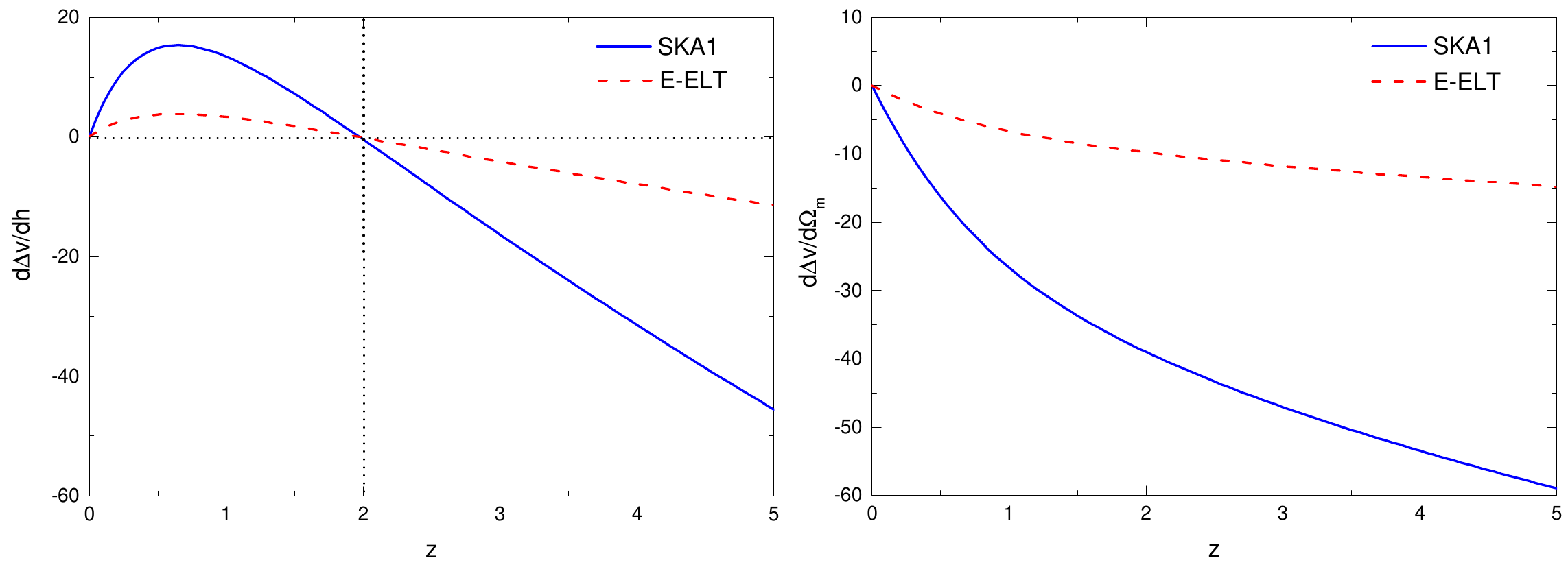}}
\caption{\label{dv-dh} {Curves of $d\Delta v/dh$ versus $z$ ({\it left}) and $d\Delta v/d\Omega_{\rm m}$ versus $z$ ({\it right}) for E-ELT and SKA in the $\Lambda$CDM model. Parameter values are set as in Fig.~\ref{mockDV}.}}
\end{figure*}

The $\chi^{2}$ function for BAO measurements is
\begin{equation}
\chi^2_{\rm BAO}=\sum\limits_{ i=1}^5 \frac{(\xi^{\rm obs}_{ i}-\xi^{\rm th}_{ i})^2}{\sigma_{ i}^2},
\end{equation}
where $\xi_{\rm th}$ and $\xi_{\rm obs}$ represent the theoretically predicted value and the experimentally measured value of the $i$-th data point for the BAO observations, respectively, and $\sigma_{i}$ is the standard deviation of the $i$-th data point.

\subsection{Redshift drift observations from E-ELT and SKA}

The actual measurement for the SL-test signal is the shift in the spectroscopic velocity ($\Delta v$) for a source in a given time interval ($\Delta t_o$). The spectroscopic velocity shift is usually expressed as \cite{Loeb:1998bu}
\begin{equation}\label{3}
   \Delta v=\frac{\Delta z}{1+z} =H_{\rm{0}} \Delta t_{o} \left[1-\frac{E(z)}{1+z}\right],
\end{equation}
where $E(z)$ is determined by a specific cosmological model.

The measurement of velocity shift will be achieved by the upcoming experiments such as the E-ELT and SKA through two different means. The E-ELT will be able to observe the Lyman-$\alpha$ absorption lines of distant quasar systems to achieve the measurement of $\Delta v$ in the redshift range of $z\in[2,5]$ \cite{Loeb:1998bu,Liske}. The SKA will measure the spectroscopic velocity shift $\Delta v$ by observing the neutral hydrogen emission signals of galaxies at the precision of one percent in the redshift range of $z\in[0,1]$. Obviously, the E-ELT and SKA experiments will be the ideal complements with each other, because of the explorations of different periods for the cosmic evolution.

\textbf{E-ELT mock data:} For the E-ELT data, as discussed in Ref.~\cite{Liske:2008ph}, the standard deviation on $\Delta v$ can be estimated as
\begin{equation}\label{4}
   \sigma_{\Delta v} = 1.35\left(\frac{2370}{S/N}\right)\left(\frac{N_{\rm QSO}}{30}\right)^{-1/2}\left(\frac{1+z_{\rm QSO}}{5}\right)^{x} ~\mathrm{cm}~\mathrm{s}^{-1},
\end{equation}
where $S/N$ is the signal-to-noise ratio of the Lyman-$\alpha$ spectrum, $N_{\rm QSO}$ is the number of observed quasars at the effective redshift $z_{\rm QSO}$, and $x$ is $1.7$ for $2\leq z\leq4$ and $0.9$ for $z\geq4$. In this work, we assume $S/N=3000$ and $N_{\rm QSO}=30$. We generate 30 mock data with a uniform distribution for the E-ELT's redshift drift observation in six redshift bins (the redshift interval $\Delta z=0.5$ for each bin), and we assume the observation time of $\Delta t_o =10$ years.

\textbf{SKA mock data:} For the case of SKA, we follow the prescription given in Refs.~\cite{Klockner:2015rqa,Martins:2016bbi} to produce the mock data of redshift drift. It is shown in Refs.~\cite{Klockner:2015rqa,Martins:2016bbi} that if SKA could have the full sensitivity and detect a billion galaxies, the evolution of the frequency shift in redshift space would be estimated to a precision of one percent. Thus, we consider the following two scenarios:
\begin{enumerate}
\item
  For SKA Phase 1, in our simulation, we produce 3 mock data of the drift $\Delta v$ in redshift $0<z<0.3$ with velocity uncertainties $\sigma_{\Delta v}$ respectively of $3\%$ in the first bin, $5\%$ in the second bin and $10\%$ in the third bin. The redshift interval $\Delta z$ is 0.1 for each bin and the timespan $\Delta t_o$ is 40 years. Note that although a timespan of 40 years is long integration time, it can be as a benchmark scenario to improve sensitivity and redshift coverage in the full SKA configuration.

\item For SKA Phase 2, we generate 10 mock data of the drift $\Delta v$ in the redshift $0<z<1$ with the velocity uncertainty $\sigma_{\Delta v}/\Delta v$ (relative error) ranging from 1\% to 10\%. Here, we adopt the same treatment method for the uncertainty as in Ref.~\cite{Martins:2016bbi}, i.e., the relative error $\sigma_{\Delta v}/\Delta v$ is assumed to be linearly increased from 1\% to 10\% in the redshift range of $z\in [0,1]$ (from low to high redshifts). To be more specific, the relative error is assumed to be 1\% in the first bin, 2\% in the second bin, and so forth. This could be reached in the timespan $\Delta t_o = 0.5$ years, which leads to an extremely competitive and ideal scenario. Note that the requirement of this scenario is $10^7$ galaxies observed in each bin~\cite{Martins:2016bbi}.
\end{enumerate}

In addition, in the mock data simulation, we adopt the scheme accordant with our previous papers \cite{Geng:2018pxk,Geng:2014ypa,Guo:2015gpa,He:2016rvp,Geng:2014hoa,Geng:2015ara,Zhang:2010im}.
In other words, the fiducial cosmology for the SL simulated data from E-ELT or SKA is chosen to be the best-fit cosmology according to the analysis of the data combination of SN+CMB+BAO in $\Lambda$CDM model, $w$CDM model, CPL model, and HDE model, respectively.

\section{Results and discussion}\label{sec:Results and Discussions}

Since the $\Lambda$CDM model is widely regarded as a prototype of the standard cosmology, we take this model as a reference model to test the constraining power of the SKA-only mock data and make an analysis of constraints on cosmological parameters when the redshift drift data of SKA and E-ELT are combined. In Fig.~\ref{mockDV}, we show the simulated redshift-drift data for E-ELT, SKA1, and SKA2, using the $\Lambda$CDM model as the fiducial model. In this figure, the curve of $\Delta v(z)$ is plotted according to Eq.~(\ref{3}), with the fiducial values of parameters given by the best fit to the SN+CMB+BAO data; the error bars on $\Delta v$, i.e., $\sigma_{\Delta v}$, for each redshift bin, are plotted according to Eq.~(\ref{4}) for E-ELT, and according to the detailed prescriptions described in the above section (the part entitled ``SKA mock data'') for SKA1 and SKA2. We find that in the E-ELT case the error of $\Delta v$ decreases with the increase of redshift, and vice versa in the SKA1 case or the SKA2 case. In Fig.~\ref{ELTvsSKA}, we plot the two-dimensional posterior contours at $68\%$ and $95\%$ confidence level (CL) in the $\Lambda$CDM model. We clearly see that using the SKA1-only mock data, the $\Lambda$CDM model can only be loosely constrained, while the model can be well constrained using the SKA2-only mock data.

In addition, form Fig.~\ref{ELTvsSKA}, we clearly see that in the $\Lambda$CDM model, from the E-ELT, $\Omega_{\rm m}$ and $h$ are in strong anti-correlation while constraints from SKA1 or SKA2 provide a positive correlation for $\Omega_{\rm m}$ and $h$, and thus the orthogonality of the two degeneracy orientations leads to a complete breaking for the parameter degeneracy. Thus, the constraints from the combination of E-ELT and SKA~(SKA1 or SKA2) would have a tremendous improvement, as shown by the gray and red contours in Fig.~\ref{ELTvsSKA}. Particularly, the result from the combination of E-ELT+SKA2 is almost as good as the constraint from the combination of SN+CMB+BAO, which implies that the redshift drift observation would have chance to be one of the most competitive cosmological probes. This may be due to the fact that the experiments of E-ELT and SKA are complementary in mapping the expansion history of the universe with a model-independent way. That is to say, these two experiments will be able to directly perform reconstruction of the expansion history of the universe in the dark matter- or dark energy-dominated epochs by using different observational techniques.

{In order to understand why $\Omega_{\rm m}$ and $h$ are in positive correlation for SKA and in anti-correlation for E-ELT, we make a deeper analysis by a comparison of the curves of derivatives of $\Delta v$ with respect to $\Omega_{\rm m}$ and $h$ versus $z$ for E-ELT and SKA. The curves of $d\Delta v/dh$ and $d\Delta v/d\Omega_{\rm m}$ versus $z$ are shown in Fig.~\ref{dv-dh}. We find that $d\Delta v/d\Omega_{\rm m}$ always decreases with increased $z$, i.e., the varying tendencies of $d\Delta v/d\Omega_{\rm m}$ for E-ELT and SKA are the same though the values of $d\Delta v/d\Omega_{\rm m}$ are not the same. However, $d\Delta v/dh$ is increasing in the most part of the SKA's redshift range of $z\in [0,1]$ (roughly the range of $z\lesssim 0.6$, covering the whole range of SKA Phase 1), and it is decreasing in the E-ELT's redshift range of $z\in [2,5]$. Meanwhile, $d\Delta v/dh$ is positive in the SKA's redshift range of $z\in [0,1]$, and it is negative in the E-ELT's redshift range of $z\in [2,5]$. The two experiments are therefore complementary in this sense, and we can clearly understand why the degeneracies between $\Omega_{\rm m}$ and $h$ can be well broken by the two experiments, as shown in Fig.~\ref{ELTvsSKA}. Notice that in Fig.~\ref{dv-dh} the blue solid curves of $d\Delta v/dh$ and $d\Delta v/d\Omega_{\rm m}$ are plotted for SKA1. We do not show the case of SKA2, since the coordinate proportions are too different, which is due to a small timespan $\Delta t_{o}$ in the SAK2 simulation.}

Meanwhile, we find that the degeneracy orientation of E-ELT+SKA1 or E-ELT+SKA2 in the parameter plane is evidently different from result for the combination of SN+CMB+BAO. This phenomenon would result in an effective breaking of the parameter degeneracy and a significant improvement of the constraints on dark energy. It is of extreme interest to know what role the redshift drift data of SKA will play in constraining dark energy in the future. Next we will explore this issue in detail.

\begin{table*}[htbp]
\tiny
\begin{center}
\caption{Priors on the free parameters for the $\Lambda$CDM, $w$CDM, CPL, and HDE models in a flat universe. }
\label{prior}
\small
\setlength\tabcolsep{9.8pt}
\renewcommand{\arraystretch}{1.2}
\begin{tabular}{cccccccccc}
\\
\hline\hline
Paramerer  &&&&&& Prior   \\ \hline

$\Omega_{\rm b}h^{2}$   &&&&&& $[0.005,0.100]$    \\
$\Omega_{\rm c}h^{2}$   &&&&&&  $[0.001,0.990]$    \\
$w_{0}$   &&&&&& $[-3.000,1.000]$    \\
$w_{a}$   &&&&&& $[-14.000,7.000]$   \\
$c$ &&&&&& $[0.200,1.200]$     \\
\hline\hline
\end{tabular}
\end{center}
\end{table*}

\begin{figure*}
\center{
\includegraphics[scale=0.4]{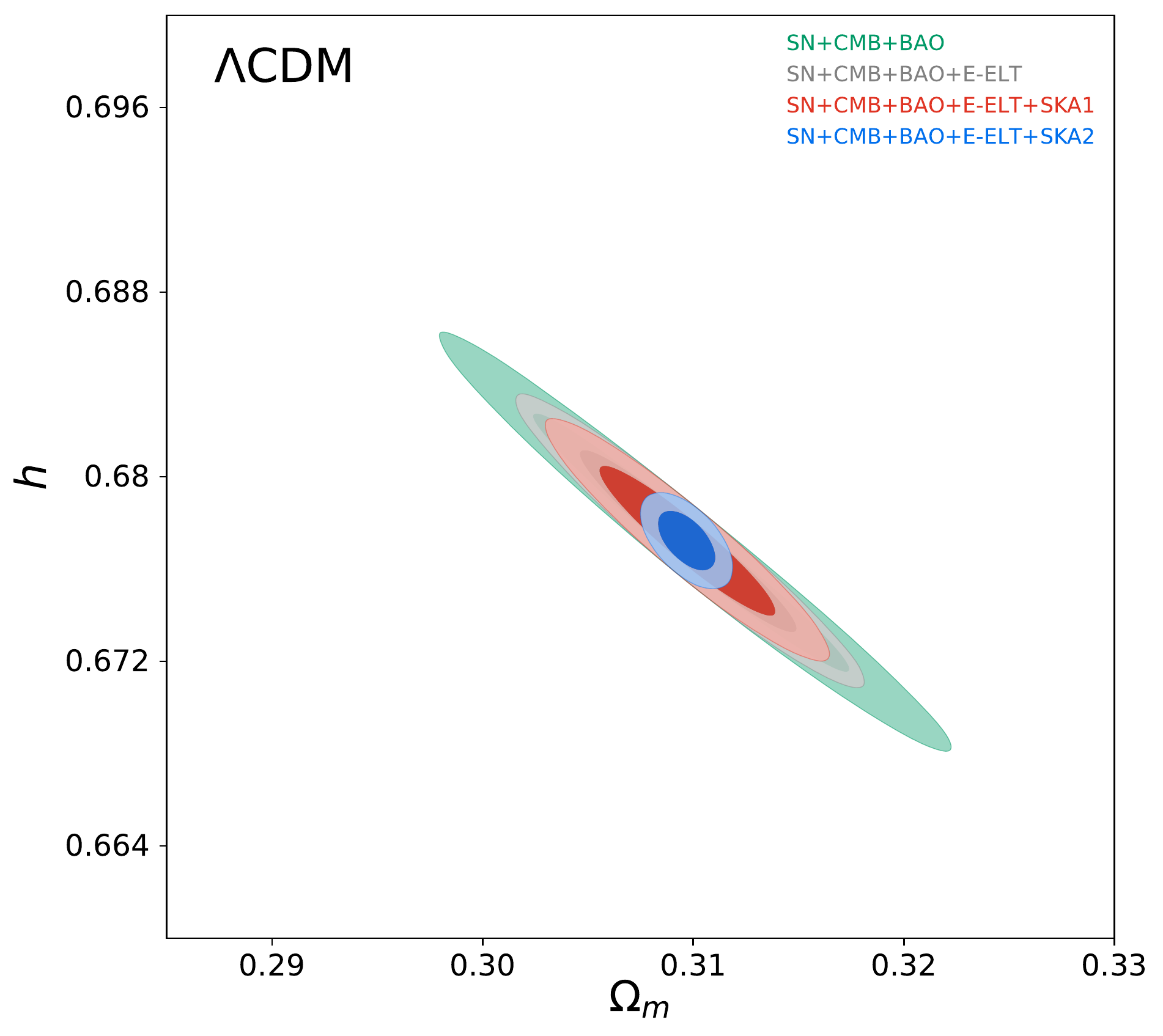}\hspace{0.4cm}
\includegraphics[scale=0.4]{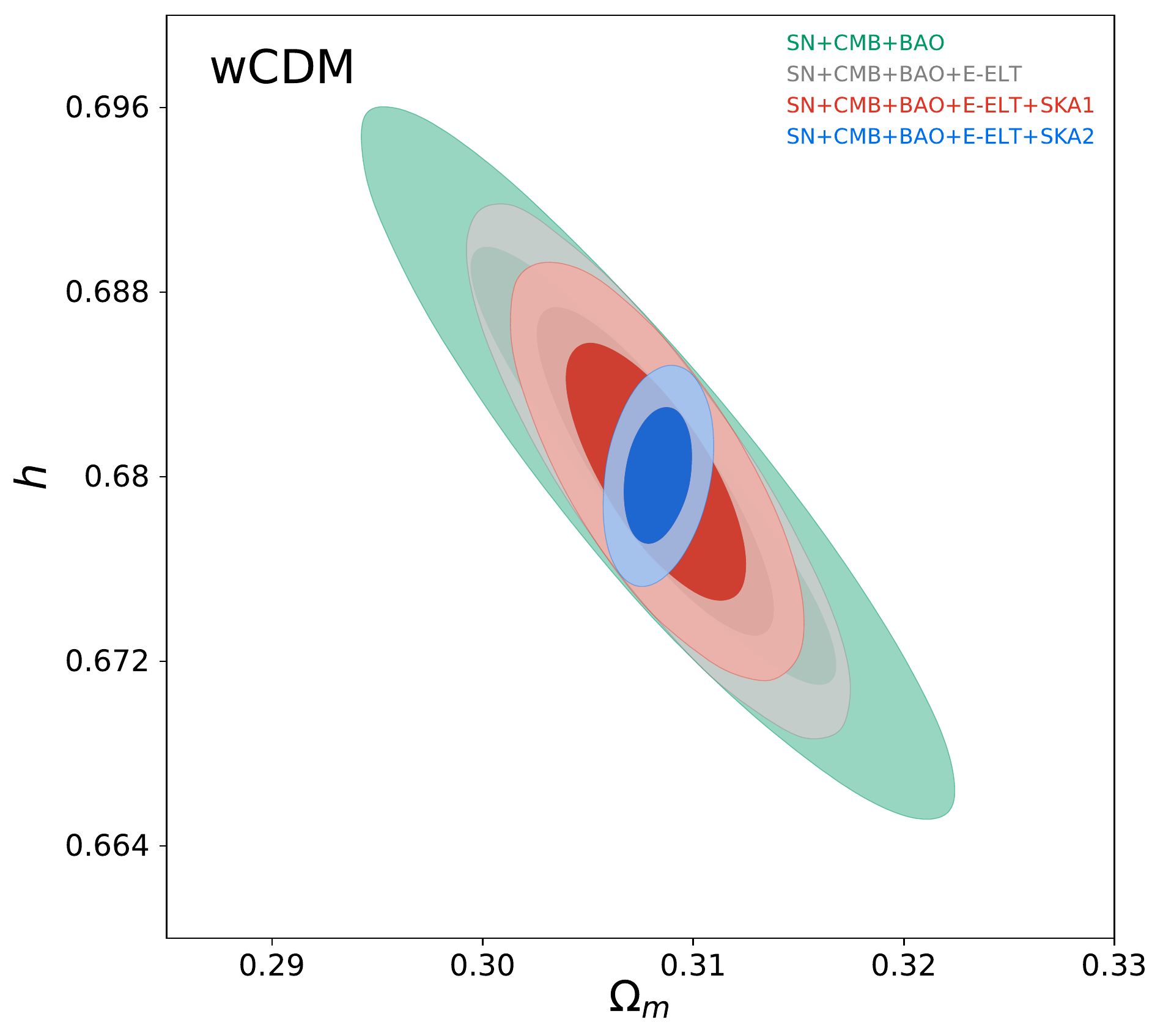}\hspace{0.4cm}
\includegraphics[scale=0.4]{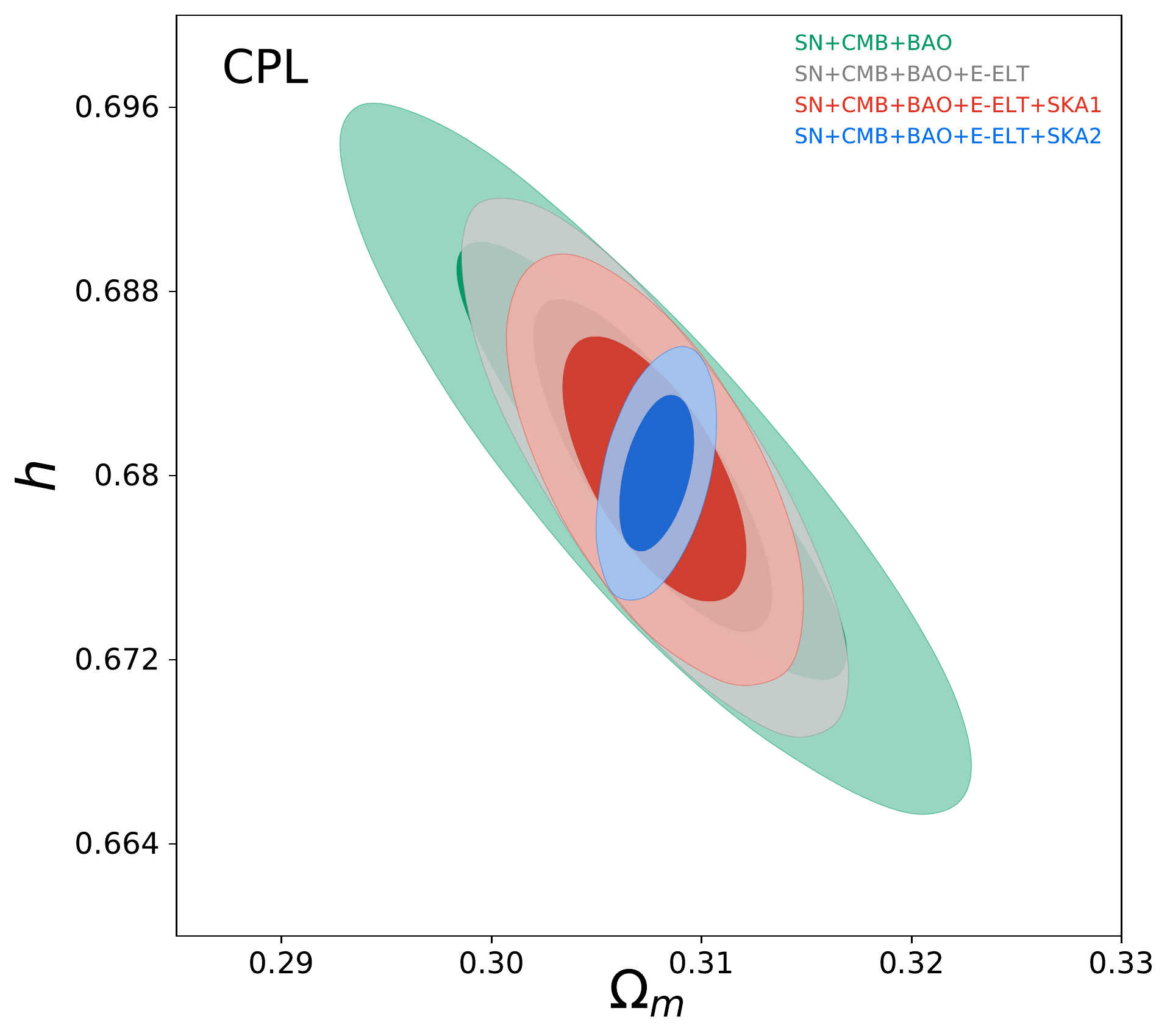}\hspace{0.4cm}
\includegraphics[scale=0.4]{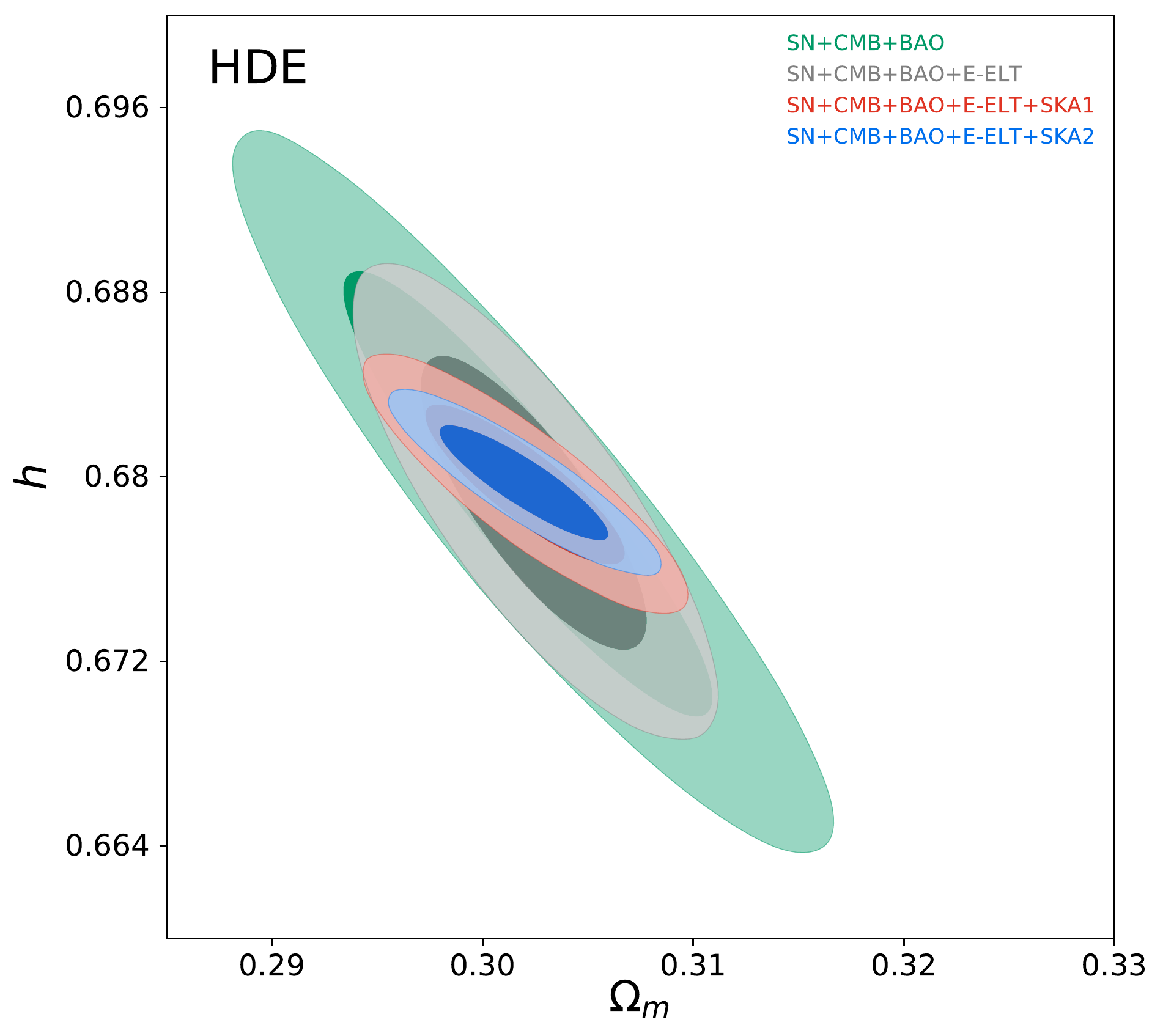}
\caption{\label{h} Constraints (1$\sigma$ and 2$\sigma$ CL) on $\Lambda$CDM, $w$CDM, CPL, and HDE models from the SN+CMB+BAO, SN+CMB+BAO+E-ELT, SN+CMB+BAO+E-ELT+SKA1, and SN+CMB+BAO+E-ELT+SKA2 data in the $\Omega_{\rm m}$--$h$ plane. }}
\end{figure*}

\begin{figure*}
\center{
\includegraphics[scale=0.4]{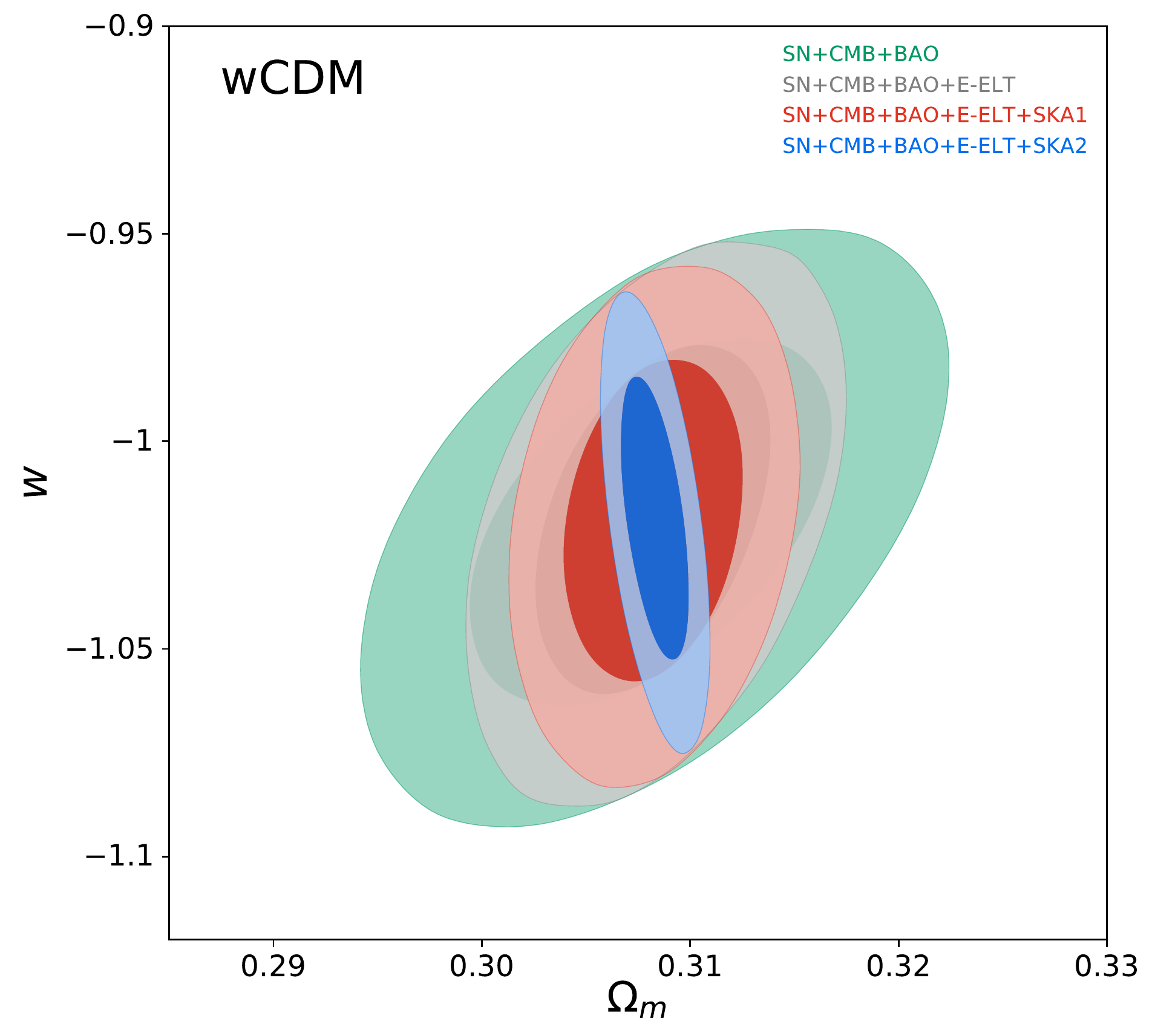}\hspace{0.4cm}
\includegraphics[scale=0.4]{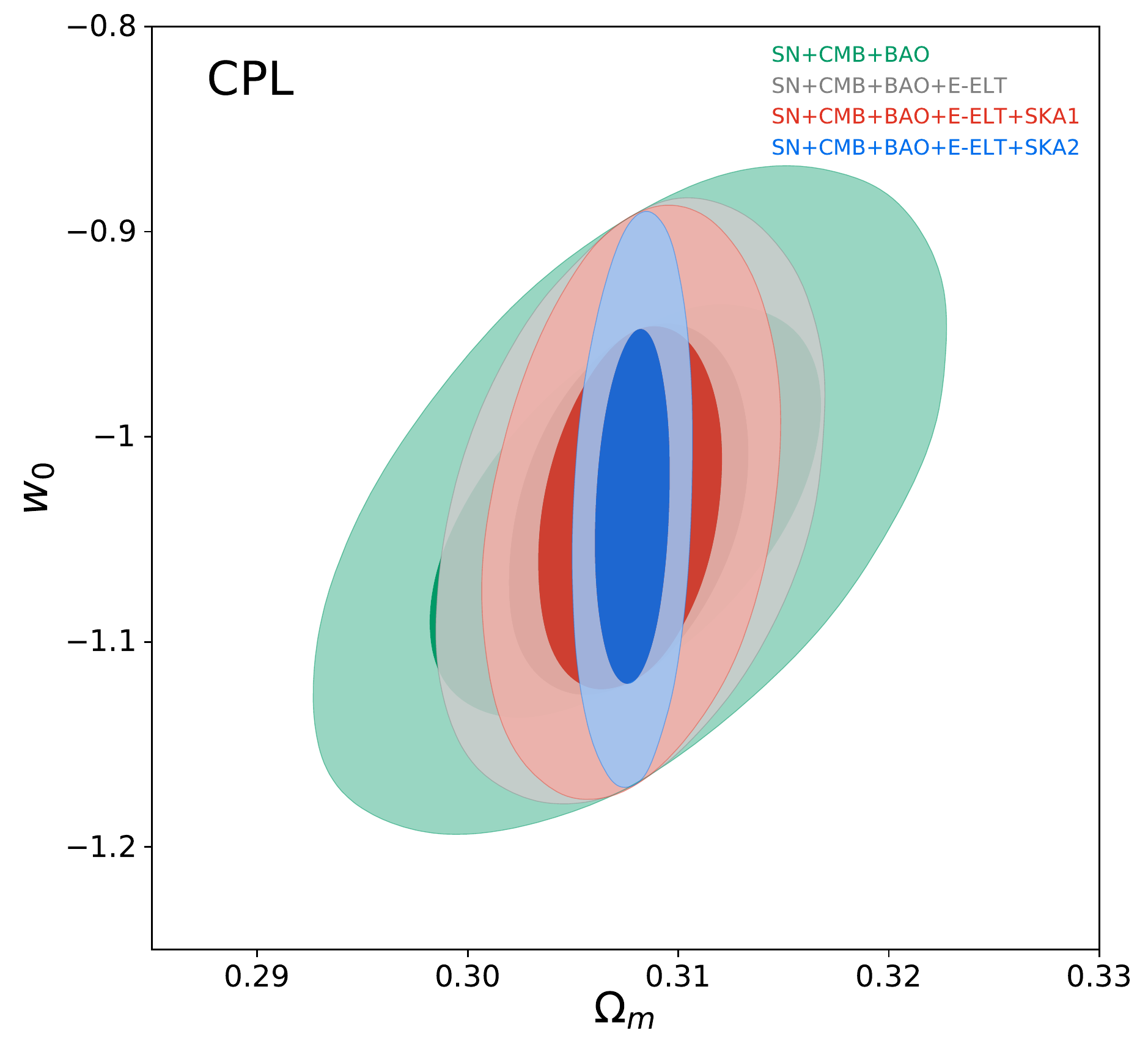}\hspace{0.4cm}
\includegraphics[scale=0.4]{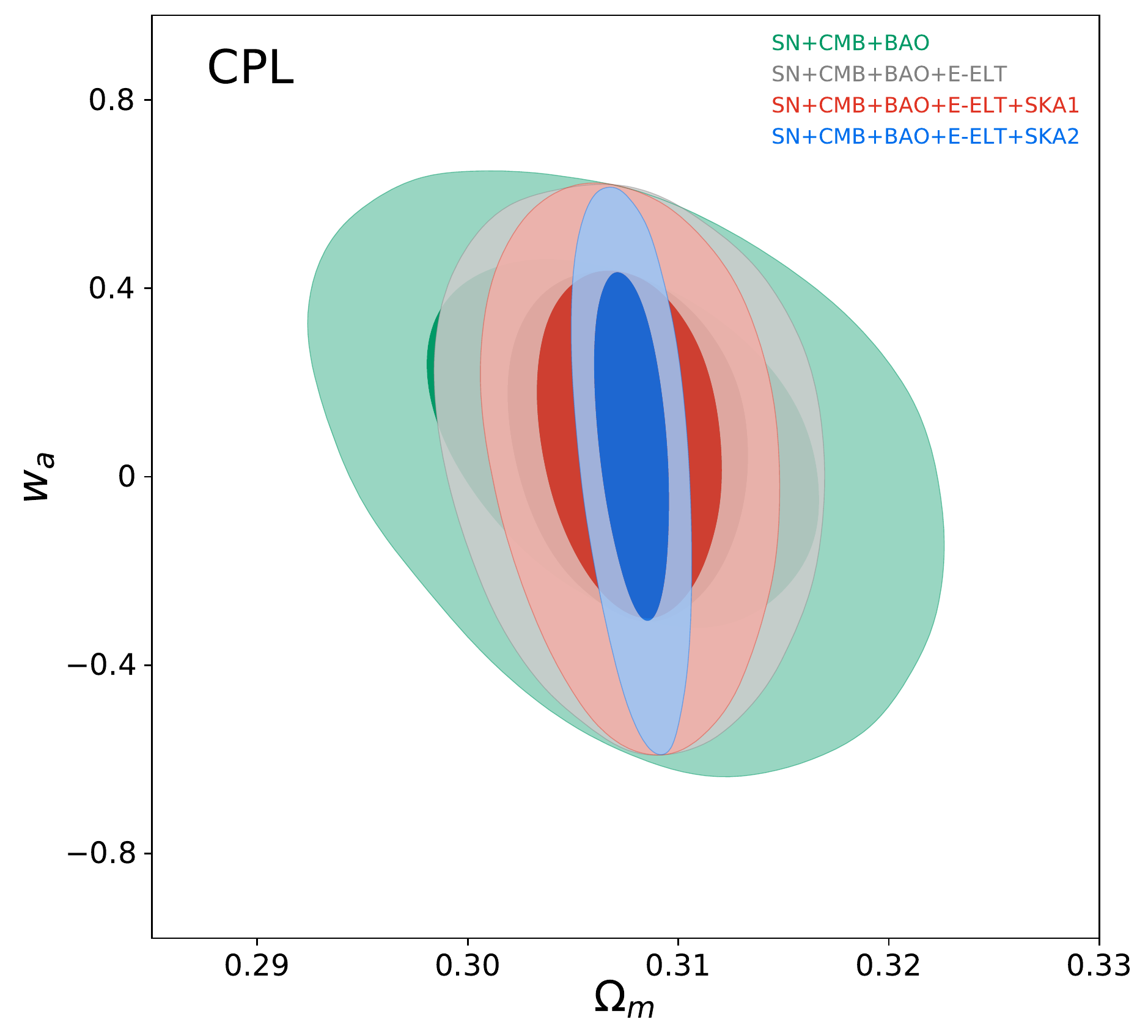}\hspace{0.4cm}
\includegraphics[scale=0.4]{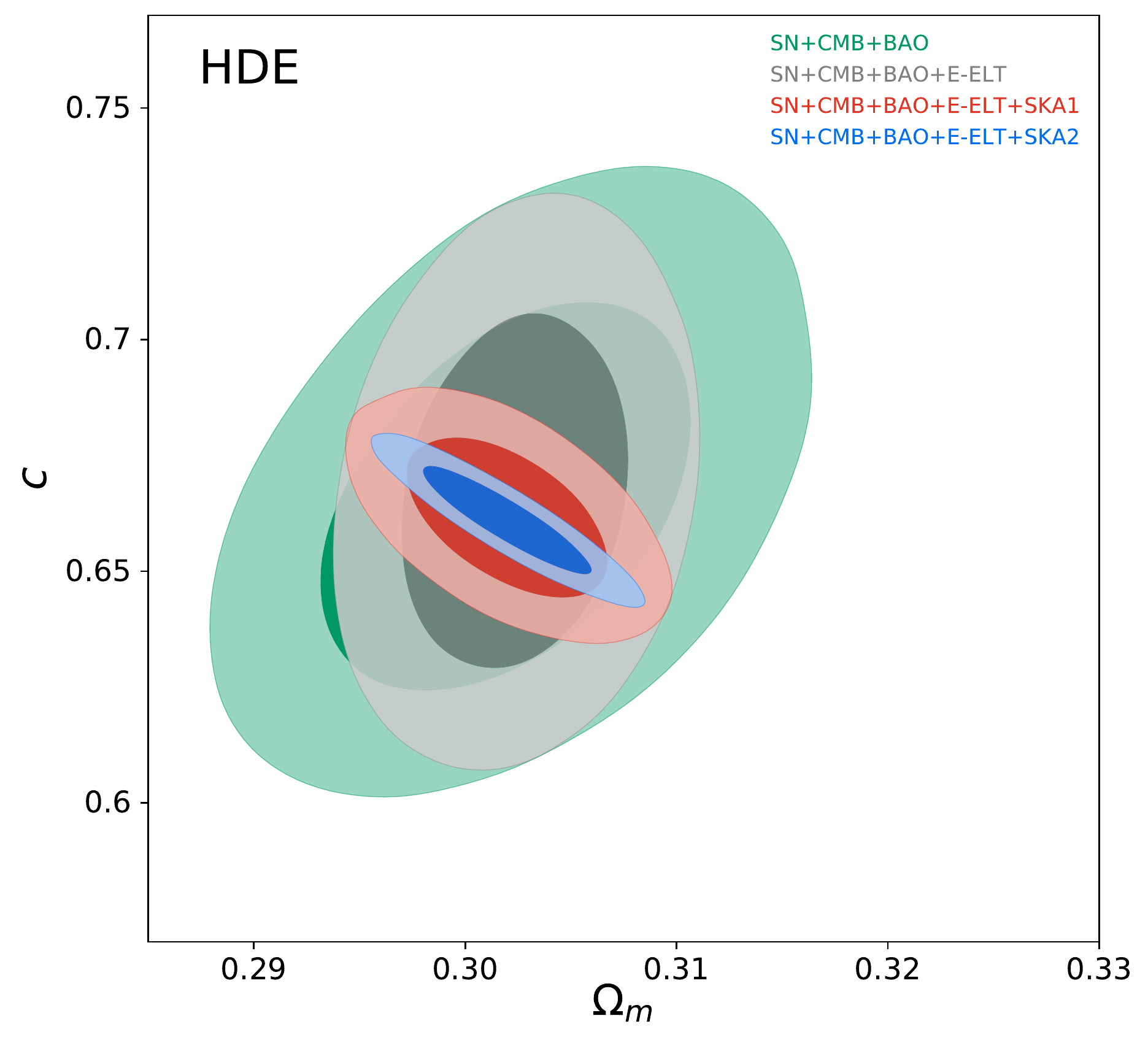}
\caption{\label{parameters} Constraints (1$\sigma$ and 2$\sigma$ CL) in the $\Omega_{\rm m}$--$w$ plane for $w$CDM model, in the $\Omega_{\rm m}$--$w_{0}$ plane for CPL model, in the $\Omega_{\rm m}$--$w_{a}$ plane for CPL model, and in the $\Omega_{\rm m}$--$c$ plane for HDE model from the SN+CMB+BAO, SN+CMB+BAO+E-ELT, SN+CMB+BAO+E-ELT+SKA1, and SN+CMB+BAO+E-ELT+SKA2 data.}}
\end{figure*}

\begin{table*}\tiny
\caption{Fitting results of parameters in the $\Lambda$CDM, $w$CDM, CPL, and HDE models using SN+CMB+BAO, SN+CMB+BAO+E-ELT, SN+CMB+BAO+E-ELT+SKA1, and SN+CMB+BAO+E-ELT+SKA2.}
\label{Parameter}
\small
\setlength\tabcolsep{1pt}
\renewcommand{\arraystretch}{1.5}
\centering
\begin{tabular}{ccccccccccccc}
\\
\hline\hline \multicolumn{1}{c}{Data} &&\multicolumn{4}{c}{SN+CMB+BAO}&&\multicolumn{4}{c}{SN+CMB+BAO+E-ELT} \\
             \cline{1-1}\cline{3-6}\cline{8-11}
Model &&$\Lambda$CDM&$w$CDM&CPL&HDE  &&$\Lambda$CDM&$w$CDM&CPL&HDE \\ \hline

$w_{0}$     && $-$
                    & $-1.0176^{+0.0395}_{-0.0435}$
                    & $-1.0416^{+0.0919}_{-0.0937}$
                    & $-$&
             & $-$
                    & $-1.0191^{+0.0391}_{-0.0391}$
                    & $-1.0373^{+0.0805}_{-0.0865}$
                    & $-$
                    \\

$w_{a}$     && $-$
                   & $-$
                   & $0.1141^{+0.3248}_{-0.3860}$
                   & $-$&
            & $-$
                   & $-$
                   & $0.0933^{+0.3289}_{-0.3482}$
                   & $-$
                   \\

$c$        && $-$
                   & $-$
                   & $-$
                   & $0.6651^{+0.0397}_{-0.0378}$&
            & $-$
                   & $-$
                   & $-$
                   & $0.6658^{+0.0363}_{-0.0344}$       \\

$\Omega_{\rm m}$ && $0.3097^{+0.0072}_{-0.0068}$
                        & $0.3082^{+0.0078}_{-0.0084}$
                        & $0.3077^{+0.0081}_{-0.0090}$
                        & $0.3018^{+0.0081}_{-0.0081}$&
                & $0.3097^{+0.0048}_{-0.0048}$
                        & $0.3080^{+0.0054}_{-0.0050}$
                        & $0.3076^{+0.0053}_{-0.0053}$
                        & $0.3023^{+0.0050}_{-0.0050}$  \\

$\emph{h}$      && $0.6772^{+0.0052}_{-0.0053}$
                        & $0.6800^{+0.0093}_{-0.0084}$
                        & $0.6798^{+0.0096}_{-0.0081}$
                        & $0.6794^{+0.0091}_{-0.0097}$&
                & $0.6771^{+0.0037}_{-0.0037}$
                        & $0.6803^{+0.0064}_{-0.0068}$
                        & $0.6800^{+0.0068}_{-0.0067}$
                        & $0.6790^{+0.0060}_{-0.0059}$\\
\hline
 \multicolumn{1}{c}{Data} &&\multicolumn{4}{c}{SN+CMB+BAO}&&\multicolumn{4}{c}{SN+CMB+BAO+E-ELT} \\
             \cline{1-1}\cline{3-6}\cline{8-11}
Model  && $\Lambda$CDM & $w$CDM & CPL & HDE&& $\Lambda$CDM & $w$CDM & CPL  & HDE \\ \hline
$w_{0}$     && $-$&
                    & $-1.0189^{+0.0361}_{-0.0358}$
                    & $-1.0396^{+0.0810}_{-0.0822}$
                    &
           & $-$
                    & $-1.0174^{+0.0309}_{-0.0324}$
                    & $-1.0395^{+0.0791}_{-0.0798}$
                    & $-$
                    \\

$w_{a}$    && $-$
                   & $-$
                   & $0.1012^{+0.3153}_{-0.3629}$
                   & $-$&
            & $-$
                   & $-$
                   & $0.0986^{+0.3206}_{-0.3515}$
                   & $-$
                   \\

$c$        && $-$
                   & $-$
                   & $-$
                   & $0.6607^{+0.0169}_{-0.0149}$&
          & $-$
                   & $-$
                   & $-$
                   & $0.6613^{+0.0105}_{-0.0110}$       \\

$\Omega_{\rm m}$    && $0.3097^{+0.0040}_{-0.0039}$
                        & $0.3083^{+0.0038}_{-0.0041}$
                        & $0.3076^{+0.0039}_{-0.0040}$
                        & $0.3019^{+0.0043}_{-0.0044}$&
               & $0.3096^{+0.0013}_{-0.0013}$
                        & $0.3083^{+0.0015}_{-0.0015}$
                        & $0.3077^{+0.0016}_{-0.0016}$
                        & $0.3018^{+0.0038}_{-0.0036}$  \\

$\emph{h}$      && $0.6772^{+0.0030}_{-0.0030}$
                        & $0.6801^{+0.0052}_{-0.0052}$
                        & $0.6799^{+0.0054}_{-0.0054}$
                        & $0.6797^{+0.0032}_{-0.0032}$&
               & $0.6772^{+0.0012}_{-0.0012}$
                        & $0.6800^{+0.0027}_{-0.0027}$
                        & $0.6798^{+0.0031}_{-0.0032}$
                        & $0.6798^{+0.0023}_{-0.0024}$ \\
\hline\hline
\end{tabular}
\end{table*}

\begin{table*}\tiny
\caption{Constraint errors of parameters in the $\Lambda$CDM, $w$CDM, CPL, and HDE models using SN+CMB+BAO, SN+CMB+BAO+E-ELT, SN+CMB+BAO+E-ELT+SKA1, and SN+CMB+BAO+E-ELT+SKA2.}
\label{Error}
\small
\setlength\tabcolsep{1pt}
\renewcommand{\arraystretch}{1.5}
\centering
\begin{tabular}{ccccccccccc}
\\
\hline\hline \multicolumn{1}{c}{Data} &&\multicolumn{4}{c}{SN+CMB+BAO}&&\multicolumn{4}{c}{SN+CMB+BAO+E-ELT} \\
             \cline{1-1}\cline{3-6}\cline{8-11}
 Model&&$\Lambda$CDM&$w$CDM&CPL&HDE  && $\Lambda$CDM &$w$CDM &CPL &HDE \\ \hline

$\sigma(w_{0})$         && $-$
                    & $0.0415$
                    & $0.0928$
                    & $-$ &
                & $-$
                    & $0.0391$
                    & $0.0835$
                    & $-$
                    \\

$\sigma(w_{a})$    && $-$
                   & $-$
                   & $0.3554$
                   & $-$&
            & $-$
                   & $-$
                   & $0.3385$
                   & $-$
                   \\

$\sigma(c)$        && $-$
                   & $-$
                   & $-$
                   & $0.0388$&
            & $-$
                   & $-$
                   & $-$
                   & $0.0353$       \\

$\sigma(\Omega_{\rm m})$ && $0.0071$
                        & $0.0081$
                        & $0.0086$
                        & $0.0081$&
                & $0.0048$
                        & $0.0052$
                        & $0.0053$
                        & $0.0050$  \\

$\sigma(\emph{h})$      && $0.0053$
                        & $0.0089$
                        & $0.0089$
                        & $0.0094$&
                & $0.0037$
                        & $0.0066$
                        & $0.0068$
                        & $0.0060$\\
\hline
 \multicolumn{1}{c}{Data} &&\multicolumn{4}{c}{SN+CMB+BAO}&&\multicolumn{4}{c}{SN+CMB+BAO+E-ELT} \\
             \cline{1-1}\cline{3-6}\cline{8-11}
Model  && $\Lambda$CDM & $w$CDM & CPL & HDE&&$\Lambda$CDM&$w$CDM&CPL& HDE \\ \hline
$\sigma(w_{0})$     && $-$
                    & $0.0360$
                    & $0.0816$
                    & $-$&
           & $-$
                    & $0.0317$
                    & $0.0795$
                    & $-$
                    \\

$\sigma(w_{a})$    && $-$
                   & $-$
                   & $0.3391$
                   & $-$&
            & $-$
                   & $-$
                   & $0.3361$
                   & $-$
                   \\

$\sigma(c)$        && $-$
                   & $-$
                   & $-$
                   & $0.0159$&
          & $-$
                   & $-$
                   & $-$
                   & $0.0108$       \\

$\sigma(\Omega_{\rm m})$    && $0.0040$
                        & $0.0040$
                        & $0.0040$
                        & $0.0044$&
               & $0.0013$
                        & $0.0015$
                        & $0.0016$
                        & $0.0037$  \\

$\sigma(\emph{h})$      && $0.0030$
                        & $0.0052$
                        & $0.0054$
                        & $0.0032$&
               & $0.0012$
                        & $0.0027$
                        & $0.0032$
                        & $0.0024$ \\
\hline\hline
\end{tabular}
\end{table*}

\begin{table*}\tiny
\caption{Constraint precisions of parameters in $\Lambda$CDM, $w$CDM, CPL, and HDE models using SN+CMB+BAO, SN+CMB+BAO+E-ELT, SN+CMB+BAO+E-ELT+SKA1, and SN+CMB+BAO+E-ELT+SKA2.}
\label{Precision}
\small
\setlength\tabcolsep{1pt}
\renewcommand{\arraystretch}{1.5}
\centering
\begin{tabular}{ccccccccccc}
\\
\hline\hline\multicolumn{1}{c}{Data} &&\multicolumn{4}{c}{SN+CMB+BAO}&&\multicolumn{4}{c}{SN+CMB+BAO+E-ELT} \\
             \cline{1-1}\cline{3-6}\cline{8-11}
Model &&$\Lambda$CDM&$w$CDM&CPL&HDE  && $\Lambda$CDM &$w$CDM &CPL &HDE \\ \hline

$\varepsilon(w_{0})$         && $-$
                    & $0.0408$
                    & $0.0891$
                    & $-$ &
                & $-$
                    & $0.0384$
                    & $0.0805$
                    & $-$
                    \\

$\varepsilon(c)$        && $-$
                   & $-$
                   & $-$
                   & $0.0583$&
            & $-$
                   & $-$
                   & $-$
                   & $0.0530$       \\

$\varepsilon(\Omega_{\rm m})$ && $0.0229$
                        & $0.0262$
                        & $0.0279$
                        & $0.0268$&
                & $0.0155$
                        & $0.0169$
                        & $0.0172$
                        & $0.0165$  \\

$\varepsilon(\emph{h})$      && $0.0078$
                        & $0.0131$
                        & $0.0130$
                        & $0.0138$&
                & $0.0054$
                        & $0.0097$
                        & $0.0100$
                        & $0.0088$\\
\hline
\multicolumn{1}{c}{Data} &&\multicolumn{4}{c}{SN+CMB+BAO}&&\multicolumn{4}{c}{SN+CMB+BAO+E-ELT} \\
             \cline{1-1}\cline{3-6}\cline{8-11}
Model && $\Lambda$CDM & $w$CDM & CPL & HDE&&$\Lambda$CDM&$w$CDM&CPL& HDE \\ \hline
$\varepsilon(w_{0})$     && $-$
                    & $0.0353$
                    & $0.0785$
                    & $-$&
           & $-$
                    & $0.0312$
                    & $0.0765$
                    & $-$
                    \\

$\varepsilon(c)$        && $-$
                   & $-$
                   & $-$
                   & $0.0241$&
          & $-$
                   & $-$
                   & $-$
                   & $0.0163$       \\

$\varepsilon(\Omega_{\rm m})$    && $0.0129$
                        & $0.0128$
                        & $0.0130$
                        & $0.0146$&
               & $0.0042$
                        & $0.0049$
                        & $0.0051$
                        & $0.0123$  \\

$\varepsilon(\emph{h})$      && $0.0044$
                        & $0.0076$
                        & $0.0079$
                        & $0.0047$&
               & $0.0018$
                        & $0.0040$
                        & $0.0047$
                        & $0.0035$ \\
\hline\hline
\end{tabular}
\end{table*}

We constrain the $\Lambda$CDM, $w$CDM, CPL and HDE models by using the data combinations of SN+CMB+BAO, SN+CMB+BAO+E-ELT, SN+CMB+BAO+E-ELT+SKA1, and SN+CMB+BAO+E-ELT+SKA2 to complete our analysis. The priors of the free parameters are given in Table \ref{prior}. Here, $\Omega_{\rm b}h^{2}$ and $\Omega_{\rm c}h^{2}$ respectively stand for the physical baryon and cold dark matter densities. The constraint results are presented in Tables~\ref{Parameter}--\ref{Precision} and Figs.~\ref{h}--\ref{parameters}. In Table~\ref{Parameter}, we show the best-fit results with the 1$\sigma$ errors quoted. The constraint errors and precisions of the cosmological parameters are given in Tables~\ref{Error}--\ref{Precision}, respectively. Here, for a parameter $\xi$, we use $\sigma(\xi)$ to denote its 1$\sigma$ error. For the cases that its distribution slightly deviates from the gaussian distribution, we adopt the value of averaging the upper-limit and lower-limit errors. We use $\varepsilon(\xi)=\sigma(\xi)/\xi_{\rm bf}$ to denote the relative error of the parameter $\xi$, where $\xi_{\rm bf}$ is its best-fit value. In this paper, for convenience, we also informally call $\varepsilon(\xi)$ the ``constraint precision'' of the parameter $\xi$. Note that we do not calculate the constraint precision for the parameter $w_a$, since its central value is close to 0. In Figs.~\ref{h}--\ref{parameters}, we show the two-dimensional posterior distribution contours of constraint results in the $\Lambda$CDM, $w$CDM, CPL and HDE models at the 68$\%$ and 95$\%$ CL.

From these figures, we clearly see that when the E-ELT mock data are combined with SN+CMB+BAO, the parameter spaces can be significantly reduced in the $\Lambda$CDM, $w$CDM, and HDE models, while there is little significant improvement in the parameter space for the CPL model. Adding the SKA1 mock data to the data combination of SN+CMB+BAO+E-ELT, the parameter spaces are sharply reduced. In particular, when the SKA2 mock data are combined with SN+CMB+BAO+E-ELT, the improvement is actually much more significant than the case of SN+CMB+BAO+E-ELT+SKA1. Meanwhile, from Fig.~\ref{parameters}, we can easily find that the E-ELT and SKA mock data can help to break the parameter degeneracies, in particular between the parameters $\Omega_{\rm m}$ and $c$ in the HDE model.

From Table~\ref{Precision}, we can easily find that the E-ELT, SKA1, and SKA2 can significantly improve the constraints on almost all the parameters to different extent, in particular for SKA2. Concretely, when the E-ELT mock data are combined with SN+CMB+BAO, the precision of $\Omega_{m}$ is improved from $2.29\%$ to $1.55\%$ in the $\Lambda$CDM model, from $2.62\%$ to $1.69\%$ in the $w$CDM model, from $2.79\%$ to $1.72\%$ in the CPL model, from $2.68\%$ to $1.65\%$ in the HDE model. The precision of $h$, $w_{0}$, and $c$ are also enhanced in the $\Lambda$CDM, $w$CDM, CPL, and HDE models; for details, see Table~\ref{Precision}. Adding the SKA1 mock data to the data combination of SN+CMB+BAO+E-ELT, the improvement of the constraint on parameter $\Omega_{\rm m}$ is from $1.55\%$ to $1.29\%$ in the $\Lambda$CDM model, from $1.69\%$ to $1.28\%$ in the $w$CDM model, from $1.72\%$ to $1.30\%$ in the CPL model, and from $1.65\%$ to $1.46\%$ in the HDE model. For the parameter $h$, the constraint is improved from $0.54\%$ to $0.44\%$ in the $\Lambda$CDM model, from $0.97\%$ to $0.76\%$ in the $w$CDM model, from $1.00\%$ to $0.79\%$ in the CPL model, and from $0.88\%$ to $0.47\%$. For the parameters of dark energy, the improvement is from $3.84\%$ to $3.53\%$ for the parameter $w$ in the $w$CDM model, from $8.05\%$ to $7.85\%$ for the parameter $w_{0}$ in the CPL model, and from $5.30\%$ to $2.41\%$ for the parameter $c$ in the HDE model.

Furthermore, when the SKA2 mock data are combined with SN+CMB+BAO+E-ELT, the improvement of the constraint on parameter $\Omega_{\rm m}$ is from $1.55\%$ to $0.42\%$ in the $\Lambda$CDM model, from $1.69\%$ to $0.49\%$ in the $w$CDM model, from $1.72\%$ to $0.51\%$ in the CPL model, and from $1.65\%$ to $1.23\%$ in the HDE model. For the parameter $h$, the constraint is improved from $0.54\%$ to $0.18\%$ in the $\Lambda$CDM model, from $0.97\%$ to $0.40\%$ in the $w$CDM model, from $1.00\%$ to $0.47\%$ in the CPL model, and from $0.88\%$ to $0.35\%$ in the HDE model. For the parameters of dark energy, the improvement is from $3.84\%$ to $3.12\%$ for the parameter $w$ in the $w$CDM model, from $8.05\%$ to $7.65\%$ for the parameter $w_{0}$ in the CPL model, and from $5.30\%$ to $1.63\%$ for the parameter $c$ in the HDE model. We also see that for the CPL model the error of $w_{a}$ is reduced by $0.74\%$ once the SKA2 data are considered. Therefore, we conclude that the redshift drift data of SKA will help to significantly improve the constraints of parameters and break the degeneracy between the parameters in constraining dark energy in the future.

\section{Conclusion}\label{sec:Conclusion}

In this work, we wish to investigate what extent the cosmological parameters can be constrained to when the redshift drift data of SKA are used and what
will happen when the combination of SKA and E-ELT mock data is considered. We use the five data sets, i.e., SKA1, SKA2, E-ELT, E-ELT+SKA1, E-ELT+SKA2, and SN+CMB+BAO to reach our aims in the $\Lambda$CDM model. We find that using the SKA2 mock data alone, the $\Lambda$CDM model can be constrained well, while the constraint is weak from the mock data of SKA1-only. When the redshift drift mock data of SKA and E-ELT are combined, the results show that the parameter space is dramatically reduced almost as good as SN+CMB+BAO. Thus, the last aim of this work is to investigate what role the redshift drift data of SKA will play in constraining dark energy in the future. To fulfill the task, we employ several concrete dark energy models, including the $\Lambda$CDM, $w$CDM, CPL, and HDE models, which are still consistent with the current observations at least to some extent.

We first use the data combination of SN+CMB+BAO to constrain the four dark energy models, and then we consider the addition of the E-ELT mock data in the data combination, i.e., we use the data combination of SN+CMB+BAO+E-ELT to constrain the models. The constraints on cosmological parameters are tremendously improved for the $\Lambda$CDM, $w$CDM, and HDE models, while E-ELT mock data do not help improve constraints in the CPL model. When adding the SKA1 mock data to the SN+CMB+BAO+E-ELT, the constraint results are significantly improved in all the four dark energy models. For example, with the help of the SKA1 mock data, the constraints on $\Omega_{\rm m}$ are improved by 10$\%$--25$\%$, and  the constraints on $h$ are improved by 15$\%$--50$\%$. Furthermore, when the SKA2 mock data are combined with the dataset of SN+CMB+BAO+E-ELT, the constraint results are tremendously improved in all the four dark energy models. Concretely, the constraints on $\Omega_{\rm m}$ are improved by 25$\%$--70$\%$, and the constraints on $h$ are improved by 50$\%$--70$\%$. We also find that the degeneracy between cosmological parameters could be effectively broken by the combination of the E-ELT and SKA mock data. Therefore, we can conclude that in the future the redshift-drift observation of SKA would help to improve the constraints in constraining dark energy and have a good potential to be one of the most competitive cosmological probes in constraining dark energy.

\begin{acknowledgments}
This work was supported by the National Natural Science Foundation of China (Grant Nos.~11875102, 11975072, 11835009, 11522540, and 11690021), the Liaoning Revitalization Talents Program (Grant No. XLYC1905011), the Fundamental Research Funds for the Central Universities (Grant No. N2005030),  and the Top-Notch Young Talents Program of China.
\end{acknowledgments}


\begin{thebibliography}{99}
\bibitem{sandage}
 A.~Sandage,
 Astrophys. J. {\bf 136}, 319 (1962).

\bibitem{Loeb:1998bu}
  A.~Loeb,
  Astrophys.\ J.\  {\bf 499}, L111 (1998)
  [astro-ph/9802122].

\bibitem{Zhang:2007zga}
  H.~B.~Zhang, W.~H.~Zhong, Z.~H.~Zhu and S.~He,
  Phys.\ Rev.\ D {\bf 76}, 123508 (2007)
  [arXiv:0705.4409 [astro-ph]].

\bibitem{Corasaniti:2007bg}
  P.~S.~Corasaniti, D.~Huterer and A.~Melchiorri,
  Phys.\ Rev.\ D {\bf 75}, 062001 (2007)
  [astro-ph/0701433].

\bibitem{Balbi:2007fx}
  A.~Balbi and C.~Quercellini,
  Mon.\ Not.\ R.\ Astron.\ Soc.\  {\bf 382}, 1623 (2007)
  [arXiv:0704.2350 [astro-ph]].

\bibitem{Liske:2008ph}
  J.~Liske {\it et al.},
  %``Cosmic dynamics in the era of Extremely Large Telescopes,''
  Mon.\ Not.\ Roy.\ Astron.\ Soc.\  {\bf 386}, 1192 (2008)
  %doi:10.1111/j.1365-2966.2008.13090.x
  [arXiv:0802.1532 [astro-ph]].

\bibitem{Zhang:2010im}
  J.~Zhang, L.~Zhang and X.~Zhang,
  %``Sandage-Loeb test for the new agegraphic and Ricci dark energy models,''
  Phys.\ Lett.\ B {\bf 691}, 11 (2010)
  %doi:10.1016/j.physletb.2010.06.013
  [arXiv:1006.1738 [astro-ph.CO]].

\bibitem{Martinelli}
  M.~Martinelli {\it et al.},
  Phys.\ Rev.\ D {\bf 86}, 123001 (2012)
  %``Probing dark energy with redshift drift,''
 [arXiv:1210.7166 [astro-ph.CO]].

\bibitem{Yuan:2013wpa}
  S.~Yuan and T.~J.~Zhang,
  %``Breaking through the high redshift bottleneck of Observational Hubble parameter Data: The Sandage-Loeb signal Scheme,''
  JCAP {\bf 1502}, no. 02, 025 (2015)
  %doi:10.1088/1475-7516/2015/02/025
  [arXiv:1311.1583 [astro-ph.CO]].

\bibitem{Zhang:2013mja}
  M.~J.~Zhang and W.~B.~Liu,
  %``Power of the redshift drift on cosmological models and expansion history,''
  arXiv:1311.6858 [astro-ph.CO].



\bibitem{Geng:2014hoa}
  J.~J.~Geng, J.~F.~Zhang and X.~Zhang,
  JCAP {\bf 1407}, 006 (2014)
  [arXiv:1404.5407 [astro-ph.CO]].

 \bibitem{Geng:2014ypa}
  J.~J.~Geng, J.~F.~Zhang and X.~Zhang,
  JCAP {\bf 1412}, no. 12, 018 (2014)
  [arXiv:1407.7123 [astro-ph.CO]].

\bibitem{Geng:2015ara}
  J.~J.~Geng, Y.~H.~Li, J.~F.~Zhang and X.~Zhang,
  %``Redshift drift exploration for interacting dark energy,''
  Eur.\ Phys.\ J.\ C {\bf 75}, no. 8, 356 (2015)
  %doi:10.1140/epjc/s10052-015-3581-8
  [arXiv:1501.03874 [astro-ph.CO]].

\bibitem{Guo:2015gpa}
  R.~Y.~Guo and X.~Zhang,
  Eur.\ Phys.\ J.\ C {\bf 76}, no. 3, 163 (2016)
  [arXiv:1512.07703 [astro-ph.CO]].

\bibitem{He:2016rvp}
  D.~Z.~He, J.~F.~Zhang and X.~Zhang,
  %``Redshift drift constraints on holographic dark energy,''
  Sci.\ China Phys.\ Mech.\ Astron.\  {\bf 60} (2017) no.3,  039511
  doi:10.1007/s11433-016-0472-1
  [arXiv:1607.05643 [astro-ph.CO]].

\bibitem{Lazkoz:2017fvx}
  R.~Lazkoz, I.~Leanizbarrutia and V.~Salzano,
  %``Forecast and analysis of the cosmological redshift drift,''
  Eur.\ Phys.\ J.\ C {\bf 78}, no. 1, 11 (2018)
 % doi:10.1140/epjc/s10052-017-5479-0
  [arXiv:1712.07555 [astro-ph.CO]].

\bibitem{Geng:2018pxk}
  J.~J.~Geng, R.~Y.~Guo, A.~Wang, J.~F.~Zhang and X.~Zhang,
  %``Prospect for cosmological parameter estimation using future Hubble parameter measurements,''
  Commun.\ Theor.\ Phys.\  {\bf 70}, no. 4, 445 (2018)
  %doi:10.1088/0253-6102/70/4/445
  [arXiv:1806.10735 [astro-ph.CO]].

\bibitem{Liu:2018kjv}
  Y.~Liu, R.~Y.~Guo, J.~F.~Zhang and X.~Zhang,
  %``Revisit of constraints on dark energy with Hubble parameter measurements including future redshift drift observations,''
  JCAP {\bf 1905}, 016 (2019)
  %doi:10.1088/1475-7516/2019/05/016
  [arXiv:1811.12131 [astro-ph.CO]].

\bibitem{Chevallier:2000qy}
  M.~Chevallier and D.~Polarski,
  %``Accelerating universes with scaling dark matter,''
  Int.\ J.\ Mod.\ Phys.\ D {\bf 10}, 213 (2001)
  %doi:10.1142/S0218271801000822
  [gr-qc/0009008].

\bibitem{Linder:2002et}
  E.~V.~Linder,
  %``Exploring the expansion history of the universe,''
  Phys.\ Rev.\ Lett.\  {\bf 90}, 091301 (2003)
 % doi:10.1103/PhysRevLett.90.091301
  [astro-ph/0208512].

\bibitem{Cohen:1998zx}
  A.~G.~Cohen, D.~B.~Kaplan and A.~E.~Nelson,
  %``Effective field theory, black holes, and the cosmological constant,''
  Phys.\ Rev.\ Lett.\  {\bf 82}, 4971 (1999)
  %doi:10.1103/PhysRevLett.82.4971
  [hep-th/9803132].

\bibitem{Li:2004rb}
  M.~Li,
  %``A Model of holographic dark energy,''
  Phys.\ Lett.\ B {\bf 603}, 1 (2004)
  %doi:10.1016/j.physletb.2004.10.014
  [hep-th/0403127].

\bibitem{Zhang:2005yz}
  X.~Zhang,
  %``Statefinder diagnostic for holographic dark energy model,''
  Int.\ J.\ Mod.\ Phys.\ D {\bf 14}, 1597 (2005)
 % doi:10.1142/S0218271805007243
  [astro-ph/0504586].

\bibitem{Zhang:2005hs}
  X.~Zhang and F.~Q.~Wu,
  %``Constraints on holographic dark energy from Type Ia supernova observations,''
  Phys.\ Rev.\ D {\bf 72}, 043524 (2005)
 % doi:10.1103/PhysRevD.72.043524
  [astro-ph/0506310].
%\cite{Zhang:2006av}

\bibitem{Zhang:2006qu}
  X.~Zhang,
  %``Dynamical vacuum energy, holographic quintom, and the reconstruction of scalar-field dark energy,''
  Phys.\ Rev.\ D {\bf 74}, 103505 (2006)
  doi:10.1103/PhysRevD.74.103505
  [astro-ph/0609699].

\bibitem{Zhang:2006av}
  X.~Zhang,
  %``Reconstructing holographic quintessence,''
  Phys.\ Lett.\ B {\bf 648}, 1 (2007)
  doi:10.1016/j.physletb.2007.02.069
  [astro-ph/0604484].

\bibitem{Zhang:2007sh}
  X.~Zhang and F.~Q.~Wu,
  %``Constraints on Holographic Dark Energy from Latest Supernovae, Galaxy Clustering, and Cosmic Microwave Background Anisotropy Observations,''
  Phys.\ Rev.\ D {\bf 76}, 023502 (2007)
 % doi:10.1103/PhysRevD.76.023502
  [astro-ph/0701405].

\bibitem{Zhang:2007es}
  J.~Zhang, X.~Zhang and H.~Liu,
  %``Holographic tachyon model,''
  Phys.\ Lett.\ B {\bf 651}, 84 (2007)
  doi:10.1016/j.physletb.2007.06.019
  [arXiv:0706.1185 [astro-ph]].
  %%CITATION = doi:10.1016/j.physletb.2007.06.019;%%
  %101 citations counted in INSPIRE as of 17 Jul 2019

%\cite{Zhang:2007an}
\bibitem{Zhang:2007an}
  J.~f.~Zhang, X.~Zhang and H.~y.~Liu,
  %``Holographic dark energy in a cyclic universe,''
  Eur.\ Phys.\ J.\ C {\bf 52}, 693 (2007)
  doi:10.1140/epjc/s10052-007-0408-2
  [arXiv:0708.3121 [hep-th]].
  %%CITATION = doi:10.1140/epjc/s10052-007-0408-2;%%
  %93 citations counted in INSPIRE as of 17 Jul 2019

\bibitem{Zhang:2008mb}
  J.~F.~Zhang, X.~Zhang and H.~Liu,
  %``Agegraphic dark energy as a quintessence,''
  Eur.\ Phys.\ J.\ C {\bf 54}, 303 (2008)
  doi:10.1140/epjc/s10052-008-0532-7
  [arXiv:0801.2809 [astro-ph]].

\bibitem{Gao:2007ep}
  C.~Gao, F.~Wu, X.~Chen and Y.~G.~Shen,
  %``A Holographic Dark Energy Model from Ricci Scalar Curvature,''
  Phys.\ Rev.\ D {\bf 79}, 043511 (2009)
  %doi:10.1103/PhysRevD.79.043511
  [arXiv:0712.1394 [astro-ph]].
  %%CITATION = doi:10.1103/PhysRevD.79.043511;%%
  %321 citations counted in INSPIRE as of 18 Jun 2019


%\cite{Zhang:2008mb}

  %%CITATION = doi:10.1140/epjc/s10052-008-0532-7;%%
  %99 citations counted in INSPIRE as of 17 Jul 2019

\bibitem{Li:2009bn}
  M.~Li, X.~D.~Li, S.~Wang and X.~Zhang,
  %``Holographic dark energy models: A comparison from the latest observational data,''
  JCAP {\bf 0906}, 036 (2009)
  %doi:10.1088/1475-7516/2009/06/036
  [arXiv:0904.0928 [astro-ph.CO]].

\bibitem{Zhang:2009un}
  X.~Zhang,
  %``Holographic Ricci dark energy: Current observational constraints, quintom feature, and the reconstruction of scalar-field dark energy,''
  Phys.\ Rev.\ D {\bf 79}, 103509 (2009)
  doi:10.1103/PhysRevD.79.103509
  [arXiv:0901.2262 [astro-ph.CO]].
  %%CITATION = doi:10.1103/PhysRevD.79.103509;%%
  %116 citations counted in INSPIRE as of 17 Jul 2019

\bibitem{Feng:2009hr}
  C.~J.~Feng and X.~Zhang,
  %``Holographic Ricci Dark Energy in Randall-Sundrum Braneworld: Avoidance of Big Rip and Steady State Future,''
  Phys.\ Lett.\ B {\bf 680}, 399 (2009)
  doi:10.1016/j.physletb.2009.09.040
  [arXiv:0904.0045 [gr-qc]].

\bibitem{Li:2009jx}
  M.~Li, X.~Li and X.~Zhang,
  %``Comparison of dark energy models: A perspective from the latest observational data,''
  Sci.\ China Phys.\ Mech.\ Astron.\  {\bf 53}, 1631 (2010)
  %doi:10.1007/s11433-010-4083-1
  [arXiv:0912.3988 [astro-ph.CO]].

\bibitem{Cui:2009ns}
  J.~L.~Cui, L.~Zhang, J.~F.~Zhang and X.~Zhang,
  %``New agegraphic dark energy as a rolling tachyon,''
  Chin.\ Phys.\ B {\bf 19}, 019802 (2010)
  doi:10.1088/1674-1056/19/1/019802
  [arXiv:0902.0716 [astro-ph.CO]].

\bibitem{Wang:2012uf}
  Y.~H.~Li, S.~Wang, X.~D.~Li and X.~Zhang,
  %``Holographic dark energy in a Universe with spatial curvature and massive neutrinos: a full Markov Chain Monte Carlo exploration,''
  JCAP {\bf 1302}, 033 (2013)
  doi:10.1088/1475-7516/2013/02/033
  [arXiv:1207.6679 [astro-ph.CO]].
  %%CITATION = doi:10.1088/1475-7516/2013/02/033;%%
  %38 citations counted in INSPIRE as of 17 Jul 2019

\bibitem{Zhang:2014ija}
  J.~F.~Zhang, M.~M.~Zhao, J.~L.~Cui and X.~Zhang,
  %``Revisiting the holographic dark energy in a non-flat universe: alternative model and cosmological parameter constraints,''
  Eur.\ Phys.\ J.\ C {\bf 74}, no. 11, 3178 (2014)
  doi:10.1140/epjc/s10052-014-3178-7
  [arXiv:1409.6078 [astro-ph.CO]].
  %%CITATION = doi:10.1140/epjc/s10052-014-3178-7;%%
  %21 citations counted in INSPIRE as of 17 Jul 2019

\bibitem{Cui:2014sma}
  J.~L.~Cui and J.~F.~Zhang,
  %``Comparing holographic dark energy models with statefinder,''
  Eur.\ Phys.\ J.\ C {\bf 74}, 2849 (2014)
 % doi:10.1140/epjc/s10052-014-2849-8
  [arXiv:1402.1829 [astro-ph.CO]].

\bibitem{Zhang:2014sqa}
  J.~F.~Zhang, J.~L.~Cui and X.~Zhang,
  %``Diagnosing holographic dark energy models with statefinder hierarchy,''
  Eur.\ Phys.\ J.\ C {\bf 74}, no. 10, 3100 (2014)
 % doi:10.1140/epjc/s10052-014-3100-3
  [arXiv:1409.6562 [astro-ph.CO]].

%\cite{Zhang:2015rha}
\bibitem{Zhang:2015rha}
  J.~F.~Zhang, M.~M.~Zhao, Y.~H.~Li and X.~Zhang,
  %``Neutrinos in the holographic dark energy model: constraints from latest measurements of expansion history and growth of structure,''
  JCAP {\bf 1504}, 038 (2015)
  doi:10.1088/1475-7516/2015/04/038
  [arXiv:1502.04028 [astro-ph.CO]].
  %%CITATION = doi:10.1088/1475-7516/2015/04/038;%%
  %33 citations counted in INSPIRE as of 17 Jul 2019

%\cite{Zhang:2015uhk}
\bibitem{Zhang:2015uhk}
  X.~Zhang,
  %``Impacts of dark energy on weighing neutrinos after Planck 2015,''
  Phys.\ Rev.\ D {\bf 93}, no. 8, 083011 (2016)
  doi:10.1103/PhysRevD.93.083011
  [arXiv:1511.02651 [astro-ph.CO]].
  %%CITATION = doi:10.1103/PhysRevD.93.083011;%%
  %39 citations counted in INSPIRE as of 17 Jul 2019


%\cite{Wang:2016tsz}
\bibitem{Wang:2016tsz}
  S.~Wang, Y.~F.~Wang, D.~M.~Xia and X.~Zhang,
  %``Impacts of dark energy on weighing neutrinos: mass hierarchies considered,''
  Phys.\ Rev.\ D {\bf 94}, no. 8, 083519 (2016)
  doi:10.1103/PhysRevD.94.083519
  [arXiv:1608.00672 [astro-ph.CO]].
  %%CITATION = doi:10.1103/PhysRevD.94.083519;%%
  %31 citations counted in INSPIRE as of 17 Jul 2019



%\cite{Zhao:2017urm}
\bibitem{Zhao:2017urm}
  M.~M.~Zhao, D.~Z.~He, J.~F.~Zhang and X.~Zhang,
  %``Search for sterile neutrinos in holographic dark energy cosmology: Reconciling Planck observation with the local measurement of the Hubble constant,''
  Phys.\ Rev.\ D {\bf 96}, no. 4, 043520 (2017)
  doi:10.1103/PhysRevD.96.043520
  [arXiv:1703.08456 [astro-ph.CO]].
  %%CITATION = doi:10.1103/PhysRevD.96.043520;%%
  %31 citations counted in INSPIRE as of 17 Jul 2019

%\cite{Feng:2018yew}
\bibitem{Feng:2018yew}
  L.~Feng, Y.~H.~Li, F.~Yu, J.~F.~Zhang and X.~Zhang,
  %``Exploring interacting holographic dark energy in a perturbed universe with parameterized post-Friedmann approach,''
  Eur.\ Phys.\ J.\ C {\bf 78}, no. 10, 865 (2018)
  doi:10.1140/epjc/s10052-018-6338-3
  [arXiv:1807.03022 [astro-ph.CO]].
  %%CITATION = doi:10.1140/epjc/s10052-018-6338-3;%%
  %5 citations counted in INSPIRE as of 17 Jul 2019








\bibitem{Zhang:2019ple}
  J.~F.~Zhang, H.~Y.~Dong, J.~Z.~Qi and X.~Zhang,
  %``Prospect for constraining holographic dark energy with gravitational wave standard sirens from the Einstein Telescope,''
  arXiv:1906.07504 [astro-ph.CO].



\bibitem{Scolnic:2017caz}
  D.~M.~Scolnic {\it et al.},
  %``The Complete Light-curve Sample of Spectroscopically Confirmed SNe Ia from Pan-STARRS1 and Cosmological Constraints from the Combined Pantheon Sample,''
  Astrophys.\ J.\  {\bf 859}, no. 2, 101 (2018)
  %doi:10.3847/1538-4357/aab9bb
  [arXiv:1710.00845 [astro-ph.CO]].

\bibitem{Tripp}
 R.~Tripp,
 %A Two-parameter luminosity correction for type Ia supernovae
 A\&A {\bf 331}, 815 (1998)

\bibitem{Guy}
SNLS Collaboration (J.~Guy {\it et al.}) A\&A {\bf 523} (2010) A7
%The Supernova Legacy Survey 3-year sample: Type Ia Supernovae photometric distances and cosmological constraints
[arXiv:1010.4743[astro-ph.CO]]

\bibitem{Ade:2015rim}
  P.~A.~R.~Ade {\it et al.} [Planck Collaboration],
  %``Planck 2015 results. XIV. Dark energy and modified gravity,''
  Astron.\ Astrophys.\  {\bf 594}, A14 (2016)
  %doi:10.1051/0004-6361/201525814
  [arXiv:1502.01590 [astro-ph.CO]].

\bibitem{Hu:1995en}
  W.~Hu and N.~Sugiyama,
  %``Small scale cosmological perturbations: An Analytic approach,''
  Astrophys.\ J.\  {\bf 471}, 542 (1996)
  %doi:10.1086/177989
  [astro-ph/9510117].

\bibitem{Alam:2016hwk}
  S.~Alam {\it et al.} [BOSS Collaboration],
  %``The clustering of galaxies in the completed SDSS-III Baryon Oscillation Spectroscopic Survey: cosmological analysis of the DR12 galaxy sample,''
  Mon.\ Not.\ Roy.\ Astron.\ Soc.\  {\bf 470}, no. 3, 2617 (2017)
  %doi:10.1093/mnras/stx721
  [arXiv:1607.03155 [astro-ph.CO]].

\bibitem{Beutler}
  F.~Beutler {\it et al.},
  Mon.\ Not.\ Roy.\ Astron.\ Soc.\  {\bf 416}, 3017 (2011)
  [arXiv:1106.3366 [astro-ph.CO]].

\bibitem{Ross:2014qpa}
  A.~J.~Ross, L.~Samushia, C.~Howlett, W.~J.~Percival, A.~Burden and M.~Manera,
  %``The clustering of the SDSS DR7 main Galaxy sample šC I. A 4 per cent distance measure at $z = 0.15$,''
  Mon.\ Not.\ Roy.\ Astron.\ Soc.\  {\bf 449}, no. 1, 835 (2015)
  %doi:10.1093/mnras/stv154
  [arXiv:1409.3242 [astro-ph.CO]].

\bibitem{Liske}
J. Liske et al.,
Top Level Requirements For ELT-HIRES, Document ESO 204697 Version 1 (2014).


\bibitem{Klockner:2015rqa}
  H.~R.~Klckner {\it et al.},
  %``Real time cosmology - A direct measure of the expansion rate of the Universe with the SKA,''
  PoS AASKA {\bf 14}, 027 (2015)
  %doi:10.22323/1.215.0027
  [arXiv:1501.03822 [astro-ph.CO]].

\bibitem{Martins:2016bbi}
  C.~J.~A.~P.~Martins, M.~Martinelli, E.~Calabrese and M.~P.~L.~P.~Ramos,
  %``Real-time cosmography with redshift derivatives,''
  Phys.\ Rev.\ D {\bf 94}, no. 4, 043001 (2016)
  %doi:10.1103/PhysRevD.94.043001
  [arXiv:1606.07261 [astro-ph.CO]].

\bibitem{Aghanim:2018eyx}
  N.~Aghanim {\it et al.} [Planck Collaboration],
  %``Planck 2018 results. VI. Cosmological parameters,''
  arXiv:1807.06209 [astro-ph.CO].

\bibitem{Xu:2016grp}
  Y.~Y.~Xu and X.~Zhang,
  %``Comparison of dark energy models after Planck 2015,''
  Eur.\ Phys.\ J.\ C {\bf 76}, no. 11, 588 (2016)
  %doi:10.1140/epjc/s10052-016-4446-5
  [arXiv:1607.06262 [astro-ph.CO]].














\end{thebibliography}
\end{document}